\pdfoutput=1
\documentclass[11pt,a4paper]{article}
\usepackage{CJKutf8}
\usepackage[utf8]{inputenc}
\usepackage{geometry}
\newgeometry{textwidth=0.8\paperwidth,top=0.05\paperheight,headheight=0.02\paperheight,headsep=0.03\paperheight,footskip=0.07\paperheight}
\usepackage{float}
\usepackage{listings}
\usepackage{caption}
\usepackage{url}
\usepackage{hyperref}
\usepackage{tikz}
\usetikzlibrary{shapes,arrows}
\usepackage[subrefformat=parens]{subcaption}
\usepackage[none]{hyphenat}
\usepackage[binary-units=true,per-mode=symbol]{siunitx}
\usepackage{textcomp}
\usepackage{adjustbox}
\usepackage{pgfplots}
\usepackage{bm}
\usepackage[bottom]{footmisc}
\usepackage{mathtools}
\usepackage{tablefootnote}
\usepackage{lineno}
\usepackage[affil-it]{authblk}
\usepackage{amssymb}
\pgfplotsset{
    tick label style = {font = {\fontsize{6 pt}{12 pt}\selectfont}},
    label style = {font = {\fontsize{8 pt}{12 pt}\selectfont}},
    legend style = {font = {\fontsize{8 pt}{12 pt}\selectfont}},  
}

\DeclareMathOperator{\E}{E}
\DeclareMathOperator{\erf}{erf}
\DeclareMathOperator{\Var}{Var}
\newcommand\acknowledgments{\section*{Acknowledgments}}
\def\secref#1{\ref{#1}~\nameref{#1}}

\title{Towards the ultimate PMT waveform analysis for neutrino and dark matter
experiments}
\date{}

\begin{document}
\author[a,b,c]{Dacheng~Xu(\begin{CJK}{UTF8}{gbsn}徐大成\end{CJK})}
\author[a,b,c,d]{Benda~Xu\footnote{Corresponding author. orv@tsinghua.edu.cn}(\begin{CJK}{UTF8}{gbsn}续本达\end{CJK})}
\author[e,f]{Erjin~Bao(\begin{CJK}{UTF8}{gbsn}宝尔金\end{CJK})}
\author[a,b,c]{Yiyang~Wu(\begin{CJK}{UTF8}{gbsn}武益阳\end{CJK})}
\author[a,b,c]{Aiqiang~Zhang(\begin{CJK}{UTF8}{gbsn}张爱强\end{CJK})}
\author[a,b,c]{Yuyi~Wang(\begin{CJK}{UTF8}{gbsn}王宇逸\end{CJK})}
\author[g]{Geliang~Zhang(\begin{CJK}{UTF8}{gbsn}张戈亮\end{CJK})}
\author[h]{Yu~Xu(\begin{CJK}{UTF8}{gbsn}徐宇\end{CJK})}
\author[a,b,c]{Ziyi~Guo(\begin{CJK}{UTF8}{gbsn}郭子溢\end{CJK})}
\author[i,2]{Jihui~Pei\footnote{Current address: School of Physics, Peking University, Beijing, China.}(\begin{CJK}{UTF8}{gbsn}裴继辉\end{CJK})}
\author[j]{Hanyang~Mao(\begin{CJK}{UTF8}{gbsn}毛晗扬\end{CJK})}
\author[j]{Jiashuo~Liu(\begin{CJK}{UTF8}{gbsn}刘家硕\end{CJK})}
\author[a,b,c]{Zhe~Wang(\begin{CJK}{UTF8}{gbsn}王\hbox{\scalebox{0.5}[1]{吉}\kern1em\scalebox{0.5}[1]{吉}}\hspace{1em}\end{CJK})}
\author[a,b,c]{Shaomin~Chen(\begin{CJK}{UTF8}{gbsn}陈少敏\end{CJK})}

\affil[a]{\footnotesize Department of Engineering Physics, Tsinghua University, Beijing, China}
\affil[b]{Center for High Energy Physics, Tsinghua University, Beijing, China}
\affil[c]{Key Laboratory of Particle \& Radiation Imaging (Tsinghua University), Ministry of Education, China}
\affil[d]{Kavli Institute for the Physics and Mathematics of the Universe, UTIAS, the University of Tokyo, Japan}
\affil[e]{National Institute of Informatics, Tokyo, Japan}
\affil[f]{Department of Informatics, The Graduate University for Advanced Studies (SOKENDAI), Tokyo, Japan}
\affil[g]{School of Securities and Futures, Southwestern University of Finance and Economics, Chengdu, China}
\affil[h]{School of Physics, Sun Yat-Sen Univesrity, Guangdong, China}
\affil[i]{Department of Physics, Tsinghua University, Beijing, China}
\affil[j]{Department of Computer Science and Technology, Tsinghua University, Beijing, China}

\maketitle

\begin{abstract}
    Photomultiplier tube~(PMT) voltage waveforms are the raw data of many neutrino and dark matter experiments. Waveform analysis is the cornerstone of data processing. We evaluate the performance of all the waveform analysis algorithms known to us and find fast stochastic matching pursuit the best in accuracy. Significant time~(up to $\times 2$) and energy~(up to $\times 1.07$) resolution boosts are attainable with fast stochastic matching pursuit, approaching theoretical limits.  Other methods also outperform the traditional threshold crossing approach in time resolution.
\end{abstract}

\tableofcontents

\section{Introduction}
\label{sec:introduction}

Waveform analysis of photomultiplier tubes~(PMT) is ubiquitous in neutrino and dark matter experiments.  Via hit time and number of photoelectrons~(PE), it provides a more accurate measurement of time and intensity of incident light on a PMT, improves event reconstruction, particle identification and definition of fiducial volume, thus promoting the physics targets.

The PMT readout system of neutrino and dark matter experiments undergoes two development stages. In Super-Kamiokande~\cite{noauthor_super-kamiokande_2003}, SNO~\cite{dunger_event_2019} and Daya Bay~\cite{daya_bay_collaboration_measurement_2017}, the time~(TDC) and charge~(QDC) to digital converters recorded the threshold crossing times and integrated charges of PMT waveforms.  Experimentalists deployed fast analog-to-digital converters to record full PMT waveforms in KamLAND~\cite{kamland_collaboration_production_2010}, Borexino~\cite{alimonti_borexino_2009}, XMASS~\cite{abe_xmass_2013} and XENON1T~\cite{xenon_collaboration_xenon1t_2019}.  This opened the flexibility of offline waveform analysis after data acquisition.  Nevertheless, limited by data volume and computational resources, early adopters emulated TDC/QDC in software with threshold and integration algorithms.  Only recently have people explored methods to extract charge and hit time of each PE~\cite{zhang_comparison_2019}.  However, there is still a great potential for improvements.  We shall see that TDC suffers from the ignorance of the second and subsequent PEs~(section~\ref{sec:time-shift-t_0}), and QDC is hindered by charge fluctuation~(section~\ref{sec:intensity-mu}).

Our mission on waveform analysis is to infer PEs from a waveform, consequently the incident light intensity over time.  The latter is the input to event reconstruction.  We shall go through all the known methods and strive towards the ultimate algorithm that retains all the available information in the data.  In section~\ref{sec:toyMC}, we discuss the principles of PE measurement in PMT-based detectors to justify the toy MC setup.  We then introduce waveform analysis algorithms and characterize their performance in section~\ref{sec:algorithm}.  Finally, in section~\ref{sec:discussion}, we discuss the impact on event reconstruction by comparing the time and intensity resolutions.

\section{Scope and Motivation}
\label{sec:toyMC}

In this section, we discuss the vital importance of waveform analysis for incident light measurements in PMT-based experiments.

\begin{figure}[!ht]
  \begin{subfigure}{0.44\textwidth}
  \includegraphics[width=\linewidth]{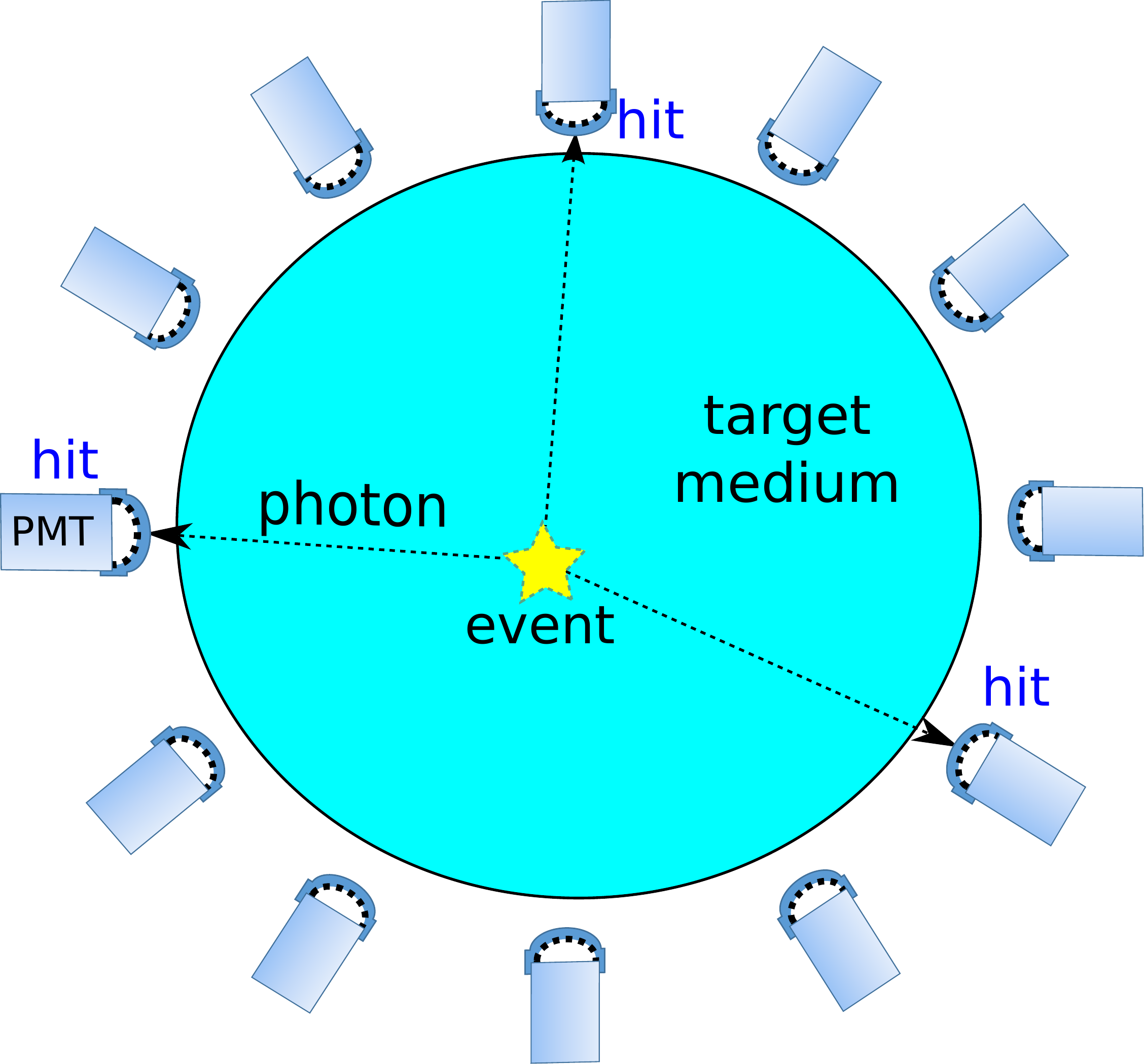}
  \caption{\label{fig:detector} Schematics of a typical PMT-based neutrino or dark matter detector.}
\end{subfigure}
\hfill
\begin{subfigure}{0.54\textwidth}
  \resizebox{\linewidth}{!}{\input{figures/profile.pgf}}
  \caption{\label{fig:time-pro} Effective light curves in the three target medium and PMT configuration settings.}
\end{subfigure}
  \caption{\subref{fig:detector} The target volume could be cylinder or polyhedra.  The PMTs' size varies from several to tens of \si{cm}.  The target medium could be pure water or ice, organic or inorganic scintillators. The principle of PMT photon counting remains the same. \subref{fig:time-pro} A scintillator paired with ultra-fast photo-sensors gives the green curve with $\tau_\ell \gg \sigma_\ell$.  A fast Cherenkov detector by the red curve has $\tau_\ell \ll \sigma_\ell$.  The blue curve combining $\tau_\ell=\SI{20}{ns}$ and $\sigma_\ell=\SI{5}{ns}$ represents a typical scintillation detector.  We can regard $\phi(t)$ as a probability density function of PE times.}
\end{figure}

Like figure~\ref{fig:detector}, a typical neutrino or dark matter detector has a large-volumed target medium surrounded by an array of PMTs. In an \textit{event}, a particle interacts with the target medium and deposits energy when passing through the detector. Part of such energy converts into visible Cherenkov or scintillation photons. A photon propagates to the boundary of the detector and converts into a PE by about \SIrange{20}{30}{\percent} quantum efficiency if it hits a PMT.

The larger the detector, the smaller the solid angle each PMT covers.  The light intensity seen by a PMT can be extremely low.  A PMT works in \textit{photon counting} or \textit{digital mode}, where it is viable to count individual PEs.  \textit{Analog mode} is the opposite, where there are so many PE pulses overlapping that the output is a continuous electric current.  Even in one event of the same detector, PMTs can operate in different modes, depending on their proximity to the event vertex.  A unified waveform algorithm should handle both extremes, and the intermediate ``overlapped, but distinguishable'' mode with an example in figure~\ref{fig:pile}.  In this study,  we cover most of the cases with PE occupancy from 0.5 to 30.

The PE, waveform and their estimators are hierarchical and intercorrelated.  We explicitly summarize the symbol conventions of this article in table~\ref{tab:symbol}.

\begin{table}[!ht]
  \centering
  \caption{definitions of symbols}
  \begin{tabular}{cll}
    \hline\hline
    variable & meaning (r.v. for random variable) & first appearance in section \\
    \hline
    $t_i, q_i$ & time and charge of the $i$-th PE (r.v.) & \secref{subsec:spe} \\
    $\bm{t}, \bm{q}$ & vector notions for sets $\{t_i\}$ and $\{q_i\}$ & \secref{sec:algorithm} \\
    $N_\mathrm{PE}$ & number of PEs (r.v.) & \secref{subsec:spe} \\
    $\hat{t}_i, \hat{q}_i$ & estimators for $t_i, q_i$ & \secref{sec:algorithm} \\
    $\hat{N}_\mathrm{PE}$ & estimator for $N_\mathrm{PE}$ & \secref{sec:time} \\
    $t'_j, N_\mathrm{s}$ & grid of PE candidate times and its size & \secref{sec:cnn} \\
    $q'_j, \bm{q}'$ & total charge at $t'_j$ and its vector (r.v.) & \secref{sec:cnn} \\
    $\alpha, \hat{\alpha}$ & scaling factor of $\bm{q}$ and its estimator & \secref{sec:shifting} \\
    $q_\mathrm{th}$ & threshold regularizer of $\bm{q}$ & \secref{sec:fourier} \\
    $z_i, \bm{z}$ & number of PEs at $t'_i$ and its vector (r.v.) & \secref{subsec:mcmc} \\
    $\bm{z}^i$ & sample of $\bm{z}$ from a Markov chain & \secref{subsec:fsmp} \\
    \hline
    $\mu$ & light intensity~(r.v.) & \secref{sec:lc} \\
    $\hat{\mu}_Q$ & charge estimator for $\mu$ & \secref{sec:intensity-mu}\\
    $t_0$ & time shift of the light curve~(r.v.) & \secref{sec:lc} \\
    $t_0^i$ & sample of $t_0$ from a Markov chain & \secref{subsec:fsmp} \\
    $\hat{t}_\mathrm{ALL}$ & ideal estimator for $t_0$ by truth $\bm{t}$ & \secref{sec:time-shift-t_0} \\
    $\hat{t}_\mathrm{1st}$ & first PE estimator for $t_0$ & \secref{sec:time-shift-t_0} \\
    $\hat{t}_\mathrm{KL}$, $\hat{\mu}_\mathrm{KL}$ & KL estimators for $t_0$ and $\mu$ & \secref{sec:pseudo} \\
    $\phi(t)$ & normalized light curve & \secref{sec:lc} \\
    $\tilde{\phi}$ & PE-sampled light curve & \secref{subsec:spe} \\
    $\phi'$ & $\sum_{j=1}^{N_\mathrm{s}}q'_j\delta(t-t'_j)$ from grid $\bm{t}'$ & \secref{sec:cnn} \\
    $\hat{\phi}$ & waveform estimator for $\tilde{\phi}$ & \secref{sec:pseudo} \\
    $\tilde{\phi}_*, \hat{\phi}_*$ & normalized $\tilde{\phi}$ and $\hat{\phi}$ & \secref{sec:W-dist} \\
    $\tilde{\Phi}(t), \hat{\Phi}(t)$ & $\int_{-\infty}^t\tilde{\phi}_*(s)\mathrm{d}s$ and $\int_{-\infty}^t\hat{\phi}_*(s)\mathrm{d}s$ & \secref{sec:W-dist}\\
    \hline
    $V_\mathrm{PE}(t)$ & shape of a single electron response & \secref{subsec:spe} \\
    $w(t), \epsilon(t)$ & PMT waveform and white noise & \secref{subsec:spe} \\
    $V_\mathrm{th}$ & threshold regularizer of $w(t)$ & \secref{sec:shifting} \\
    $\bm{w}$ & vector notion of discretized $w(t)$ & \secref{subsec:fsmp} \\
    $\tilde{w}$ & smoothed $w$, approximating $w - \epsilon$ & \secref{sec:fourier} \\
    $\tilde{w}_*$, $V_{\mathrm{PE}*}$ & normalized $\tilde{w}$ and $V_\mathrm{PE}$ & \secref{sec:lucyddm} \\
    $\hat{w}$ & $\hat{\phi} \otimes V_\mathrm{PE}$ for estimating $w$ & \secref{sec:rss} \\
    $w'$ & $\phi' \otimes V_\mathrm{PE}$ from grid $\bm{t}'$ & \secref{sec:regression} \\
    \hline\hline
  \end{tabular}
  \label{tab:symbol}
\end{table}
\subsection{Light curve}
\label{sec:lc}
The \textit{light curve} is the time evolution of light intensity illuminating a PMT,
\begin{equation}
  \label{eq:light-curve}
  \mu\phi(t-t_0)
\end{equation}
where $\mu$ is the intensity factor, $t_0$ is the time shift factor, and $\phi(\cdot)$ is the normalized shape function. For simplicity, we parameterize the scintillation light curve as an exponential distribution and the Cherenkov one by a Dirac delta function.  It is convenient to model the PMT transit time spread~(TTS) in $\phi(t)$ as a Gaussian smear, giving an \textit{ex-Gaussian} or \textit{exponentially modified Gaussian}~\cite{li_separation_2016},
\begin{align}
    \phi(t) = \frac{1}{2\tau_\ell} \exp\left(\frac{\sigma_\ell^2}{2\tau_\ell^2}-\frac{t}{\tau_\ell}\right) \left[1 - \erf\left( \frac{\sigma_\ell}{\sqrt{2}\tau_\ell} - \frac{t}{\sqrt{2}\sigma_\ell} \right)\right],
    \label{eq:time-pro}
\end{align}
where subscript $\ell$ stands for ``light curve'' and $\sigma_\ell$ encodes the timing uncertainty mainly from TTS. $\phi(t)$ of Cherenkov light is a pure Gaussian by taking $\tau_\ell \rightarrow 0$. Figure~\ref{fig:time-pro} illustrates 3 examples of $\phi(t)$. 

\subsection{Single electron response}
\label{subsec:spe}

A PE induced by a photon at the PMT photocathode is accelerated, collected, and amplified by several stages into $\num[retain-unity-mantissa=false]{\sim 1e7}$ electrons, forming a voltage pulse $V_\mathrm{PE}(t)$ in the PMT output.  Wright~et~al.~\cite{wright_low_1954} formulated the cascade multiplication of secondary electrons assuming the amplification of each stage following Poisson distribution.  Breitenberger~\cite{breitenberger_scintillation_1955} compared the statistical model with a summary of laboratory measurements observing the gain variance to be larger than predicted. Percott~\cite{prescott_statistical_1966} used Polya distribution to account for the extra variance of Poisson rate non-uniformity.  With modern high gain PMTs ($\num[retain-unity-mantissa=false]{\sim 1e7}$), Caldwell et al.~\cite{caldwell_characterization_2013} from MiniCLEAN and Amaudruz et al.~\cite{amaudruz_-situ_2019} from DEAP-3600 suggested gamma distribution as the continuous counterpart of the Polya.  Neves et al.~\cite{neves_calibration_2010} from ZEPLIN-III did a survey of the literature but prefer to model the gain in a data-driven way by calibration, without assuming any well-known probability distributions.  We choose gamma distribution in this work over Gaussian because the amplification is always positive.

A sample of PEs from the light curve $\phi(t)$ in eq.~\eqref{eq:time-pro} can be formulated as several delta functions, also known as sparse spike train~\cite{levy_reconstruction_1981}, 
\begin{equation}
  \label{eq:lc-sample}
  \tilde{\phi}(t) = \sum_{i=1}^{N_{\mathrm{PE}}} q_i \delta(t-t_i),
\end{equation}
where $N_\mathrm{PE}$ is the number of PEs following Poisson distribution with parameter $\mu$.  $t_i$ is the hit time of the $i$-th PE, $q_i$ is the relative charge of the $i$-th PE from gamma distribution.  We set the shape ($k=1/0.4^2$) and scale ($\theta=0.4^2$) parameters of gamma so that $\E[q_i] = 1$ and $\Var[q_i] = 0.4^2$, corresponding to a typical first-stage amplification of 7--8\footnote{For large gain and Poisson-distributed first-stage amplicication $M_1$, $\Var[q_i] \approx \frac{1}{M_1-1}$.}.

\begin{figure}[H]
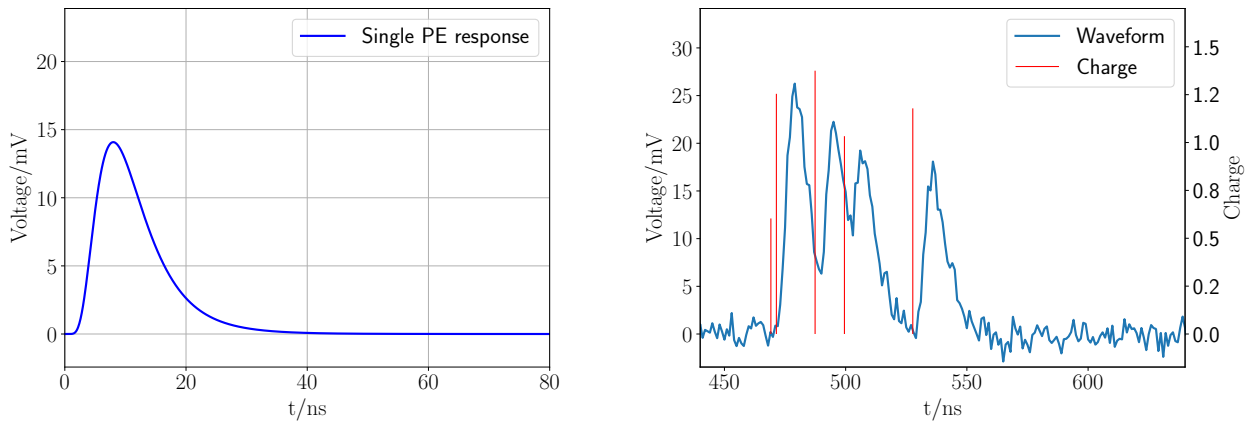

  \begin{subfigure}{.49\textwidth}
    \centering
    \resizebox{\textwidth}{!}{\input{figures/spe.pgf}}
    \caption{\label{fig:spe} Single PE response $V_\mathrm{PE}(t)$ in eq.~\eqref{eq:dayaspe}.}
  \end{subfigure}
  \begin{subfigure}{.49\textwidth}
    \centering
    \resizebox{\textwidth}{!}{\input{figures/wave.pgf}}
    \caption{\label{fig:pile} PE pile-up and white noise in an PMT waveform.}
  \end{subfigure}
  \caption{A single PE from a PMT induces a voltage pulse in \subref{fig:spe}. Multiple PEs pile up at a PMT form an input waveform $w(t)$ in \subref{fig:pile}, when PEs are barely separable from each other visually. We shall discuss the output charges $\hat{\bm{t}}, \hat{\bm{q}}$ of \subref{fig:pile} in section~\ref{sec:algorithm}. }
\end{figure}

Birks~\cite{birks_theory_1967} summarized laboratory measurements of single-electron-response~(SER) pulse shape and indicated a crude assumption of Gaussian shape is adequate, but also mentioned asymmetric model of $t e^{-bt}$ having a faster rise than decay by Hamilton and Wright~\cite{hamilton_transit_1956}.  To model the rising edge curvature better than Hamilton and Wright, S.~Jetter~et al.~\cite{jetter_pmt_2012} from DayaBay used log-normal as a convenient phenomenological representation of the SER pulse.  As in figure~\ref{fig:spe}, the smooth rising curve of log-normal fits measurements better and the falling component captures the exponential decay characteristics of RC circuit in the electronics readout.  Caldwell~et~al.~\cite{caldwell_characterization_2013}, Caravaca et al.~\cite{caravaca_experiment_2017} from CHESS and Kaptanoglu~\cite{kaptanoglu_characterization_2018} used the same parameterization and found a reasonable match with experimental data.  A better model embracing the underlying physics mechanism may be developed in the future, but for this waveform analysis study, we adopt the log-normal SER pulse as eq.~\eqref{eq:dayaspe} without loss of generality.
\begin{equation}
  V_\mathrm{PE}(t) = V_{0}\exp\left[-\frac{1}{2}\left(\frac{\log(t/\tau_\mathrm{PE})}{\sigma_\mathrm{PE}}\right)^{2}\right],
  \label{eq:dayaspe}
\end{equation}
where shape parameters $\tau_\mathrm{PE}=\SI{8}{ns}$, $\sigma_\mathrm{PE}=\SI{0.5}{ns}$ and $V_{0}=\SI{14.08}{mV}$, see figure~\ref{fig:spe}.

A noise-free waveform $\tilde{w}(t)$ is a convolution of $\tilde{\phi}(t)$ and $V_\mathrm{PE}(t)$, and the PMT voltage output waveform $w(t)$ is a time series modeled by the sum of $\tilde{w}(t)$ and a Gaussian white noise $\epsilon(t)$,
\begin{equation}
  \label{eq:1}
  \begin{aligned}
    \tilde{w}(t) &= \tilde{\phi}(t) \otimes V_\mathrm{PE}(t) \\
    w(t) &= \tilde{w}(t) + \epsilon(t) = \sum_{i=1}^{N_\mathrm{PE}} q_i V_\mathrm{PE}(t-t_i) + \epsilon(t).
  \end{aligned}
\end{equation}
See figure~\ref{fig:pile} for an example.

We do not dive into pedestals or saturation for simplicity.  We also assume the SER pulse $V_\mathrm{PE}(t)$, the variance of charge $\Var[q_i]$ and the distribution of noise $\epsilon(t)$ are known.  Otherwise, they should be measured by PMT calibrations and modeled with uncertainty.

\subsection{Measurement of incident light}
\label{sec:time}
We see in figure~\ref{fig:pile} that pile-ups and noises hinder the time $t_i$ and charge $q_i$ of the PEs. Fortunately, event reconstruction only takes the time shift $t_0$ and the intensity $\mu$ in eq.~\eqref{eq:light-curve} as inputs, where $t_0$ carries the time of flight information and $\mu$ is the expected $N_\mathrm{PE}$ in a real detector. The former directly translates into position resolution by multiplying speed-of-light, while the latter dominates energy resolution.  All the uncertainties of $\hat{t}_i$, $\hat{q}_i$ and $\hat{N}_\mathrm{PE}$ are reflected in $\hat{t}_0$ and $\hat{\mu}$. Classical TDC extracts the waveform's threshold crossing time $\hat{t}_\mathrm{1st}$ to approximate the hit time of the first PE, while QDC extracts total charge $Q$ from waveform integration to estimate $\mu$ by $\hat{\mu}_Q$.

\subsubsection{Time $t_0$}
\label{sec:time-shift-t_0}

$\hat{t}_\mathrm{1st}$ is a biased estimator of $t_0$.  It is affected by the light intensity $\mu$: the larger the $\mu$, the more negative bias $\hat{t}_\mathrm{1st}$ has.  We define the resolution of $\hat{t}_\mathrm{1st}$ as the standard deviation of its bias $\sqrt{\Var[\hat{t}_\mathrm{1st} - t_0]}$. From a hypothetical perfect measurement of $t_i$, we define an ideal maximum likelihood estimator~(MLE) $\hat{t}_\mathrm{ALL}$ to capture time information of all the PEs,
\begin{equation}
  \label{eq:2}
  \hat{t}_\mathrm{ALL} = \arg\underset{t_0}{\max} \prod_{i=1}^{N_\mathrm{PE}} \phi(t_i-t_0).
\end{equation}
The corresponding resolution $\sqrt{\Var[\hat{t}_\mathrm{ALL} - t_0]}$ serves as the reference for method evaluation.

\begin{figure}[H]
  \begin{subfigure}{.49\textwidth}
    \centering
    \resizebox{\textwidth}{!}{\input{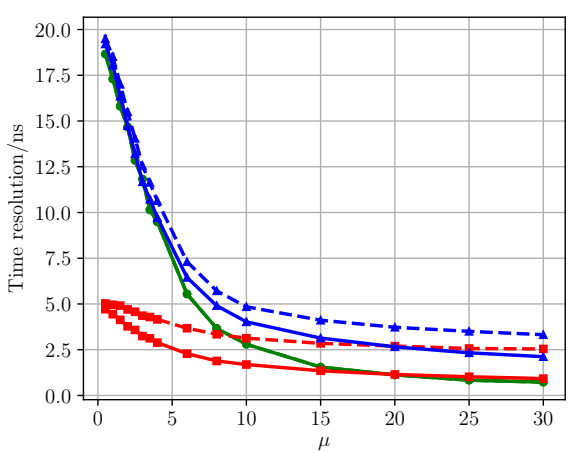}}
    \caption{\label{fig:reso-diff} Absolute time resolutions.}
  \end{subfigure}
  \begin{subfigure}{.49\textwidth}
    \centering
    \resizebox{\textwidth}{!}{\input{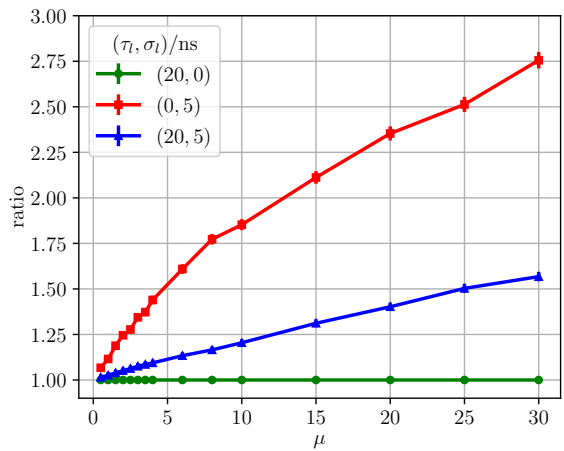}}
    \caption{\label{fig:reso-diff-r} Relative time resolutions $\sqrt{\frac{\Var[\hat{t}_{\mathrm{1st}} - t_0]}{\Var[\hat{t}_\mathrm{ALL} - t_0]}}$.}
  \end{subfigure}
  \caption{Time resolution comparisons between $\hat{t}_{\mathrm{ALL}}$~(solid lines, when using the information of all PEs) and $\hat{t}_\mathrm{1st}$~(dashed lines, using the first PE instead of the threshold crossing time to eliminate the influence from TDC).  The difference is manifested especially when $\sigma_\ell$ and $\mu$ are large. The three colors are different cases of $(\tau_\ell, \sigma_\ell)/\si{ns}$ representing the scintillation with ultra-fast PMTs~$(20, 0)$, Cherenkov~$(0, 5)$ and scintillation coupled with PMT TTS~$(20, 5)$.}
\end{figure}

To characterize the difference between $\hat{t}_\mathrm{1st}$ and $\hat{t}_\mathrm{ALL}$, we scan $\mu$ from \numrange{0}{30} for each light curve in figure~\ref{fig:time-pro}. We generate a sample of $\num[retain-unity-mantissa=false]{1e4}$ waveforms having at least 1 PE for every triplet of $(\tau_\ell, \sigma_\ell, \mu)$.  Figure~\ref{fig:reso-diff} shows a substantial difference between the two estimators, only except two cases: when $\tau_\ell \gg \sigma_\ell$, because $\hat{t}_\mathrm{ALL}$ reduces to $\hat{t}_\mathrm{1st}$(=$\min_i t_i$) for an exponential light curve; when $\mu \to 0$, because at most 1 PE is available.  

In general for $\sigma_\ell > 0$ and $\mu >0$, we notice that TDC or equivalent algorithms to $\hat{t}_\mathrm{1st}$ imposes significant resolution loss.  For Cherenkov and scintillation experiments with non-negligible PMT TTS and occupancy, we shall explore more sophisticated waveform analysis algorithms to go beyond $\hat{t}_{\mathrm{1st}}$ and recover the accuracy of $\hat{t}_\mathrm{ALL}$ in eq.~\eqref{eq:2} from waveform in eq.~\eqref{eq:1}.

\subsubsection{Intensity $\mu$}
\label{sec:intensity-mu}

The classical way is to measure light intensity $\mu$ by integration. Noting $Q_0 = {\int V_\mathrm{PE}(t) \mathrm{d} t}$, the charge estimator $\hat{\mu}_Q$ for QDC is
\begin{equation}
  \begin{aligned}
  \label{eq:mu-q}
  \hat{\mu}_Q = \frac{1}{Q_0}\int w(t) \mathrm{d} t &= \frac{1}{Q_0}\int \left[\sum_{i=1}^{N_\mathrm{PE}} q_i V_\mathrm{PE}(t-t_i) + \epsilon(t) \right] \mathrm{d} t \\
  & = \sum_{i=1}^{N_\mathrm{PE}} q_i + \frac{1}{Q_0} \int \epsilon(t) \mathrm{d} t.
  \end{aligned}
\end{equation}
Its expectation and variance are,
\begin{equation}
  \label{eq:energy}
  \begin{aligned}
    \E[\hat{\mu}_Q] &= \E[N_\mathrm{PE}] \\
    \Var[\hat{\mu}_Q] &= \Var\left[\sum_{i=1}^{N_\mathrm{PE}} q_i\right] + \Var\left[\frac{1}{Q_0} \int \epsilon(t) \mathrm{d} t\right] \\
    &= \E[q^2] \Var[N_\mathrm{PE}] + \frac{T}{Q_0^2} \Var[\epsilon] \\
    &= (1 + \Var[q]) \Var[N_\mathrm{PE}] + \frac{T}{Q_0^2} \Var[\epsilon]. \\
  \end{aligned}
\end{equation}
where the first term of the variance is from a compound Poisson distribution with a gamma jump, and $T$ in the second one is a constant proportional to the time window.  Carefully lowering $T$ could reduce the disturbance of $\Var[\epsilon]$. The resolution of $\hat{\mu}_Q$ is affected by the Poisson fluctuation of $N_\mathrm{PE}$, the charge resolution of a PE $\sigma_\mathrm{q}$ and the white noise $\sigma_\epsilon$.

Sometimes we mitigate the impact of $\Var[q]$ and $\Var[\epsilon]$ by rounding $\hat{\mu}_Q$ to integers.  It works well for $N_\mathrm{PE} \le 1$, which is equivalently a hit-based 0-1 $\mu$ estimator $\hat{\mu}_\mathrm{hit}$.  But for $N_\mathrm{PE} \ge 2$, it is hard to interpret $\hat{\mu}_Q$ rounding by physics principles and $\hat{\mu}_\mathrm{hit}$ does not gain any additional information from the extra PEs.

The goal of waveform analysis is to eliminate the impact from $\Var[q]$ and $\Var[\epsilon]$ as much as possible.  The pure Poisson fluctuation of true PE counts $\sqrt{\Var[N_\mathrm{PE}]} = \sqrt{\mu}$ is the resolution lower bound and reference to $\mu$ estimators.

\subsubsection{Shape $\phi(\cdot)$}
\label{sec:shape-phicdot}

The shape of a light curve is determined by light emission time profile, PMT timing and light propagation, including refraction, reflection, dispersion and scattering.  $\phi(\cdot)$ thus depends on event locations.  In this article, we model $\phi(\cdot)$ by eq.~\eqref{eq:time-pro} for simplicity and leave the $\phi(\cdot)$ variations to event reconstruction in future publications.
\section{Algorithms and their performance}
\label{sec:algorithm}

Waveform analysis is to obtain $t_i$ and $q_i$ estimators $\hat{t}_i$ and $\hat{q}_i$ from waveform $w(t)$, where the output indices $i$ are from 1 to $\hat{N}_\mathrm{PE}$ and $\hat{N}_\mathrm{PE}$, an estimator of $N_\mathrm{PE}$ in eq.~\eqref{eq:lc-sample}. Figure~\ref{fig:pile} illustrates the input waveform $w(t)$ and the outputs charge $\bm{\hat{t}}, \hat{\bm{q}}$ obtained from $w(t)$, where boldface $\hat{\bm{t}}$ denotes the vector $\hat{t}_i$. 

$\hat{N}_\mathrm{PE}$ may fail to estimate $N_\mathrm{PE}$ due to the fluctuation of $q_i$ and the ambiguity of $\hat{t}_i$. For example, 1, 2 and even 3~PEs can generate the same charge as $1.6$ units.  A single PE charged $1$ might be misinterpreted as 2~PEs at consecutive $\hat{t}_i$ and $\hat{t}_{i+1}$ with $\hat{q}_i=\hat{q}_{i+1}=0.5$.

\subsection{Evaluation criteria}
\label{sec:criteria}
Subject to such ambiguity of $t_i/q_i$, we introduce a set of evaluation criteria to assess the algorithms' performance.

\subsubsection{Kullback-Leibler divergence}
\label{sec:pseudo}

We construct a light curve estimator $\hat{\phi}(t)$ from $\bm{\hat{t}}$, $\bm{\hat{q}}$ and $\hat{N}_\mathrm{PE}$,
\begin{equation}
  \label{eq:lc}
  \hat{\phi}(t) = \sum_{i=1}^{\hat{N}_\mathrm{PE}} \hat{q}_i\delta(t-\hat{t}_i),
\end{equation}
which resembles eq.~\eqref{eq:lc-sample}.

Basu et al.'s \textit{density power divergence}~\cite{basu_robust_1998} contains the classical Kullback-Leibler~(KL) divergence~\cite{kullback_information_1951} as a special case.  Non-normalized KL divergence is defined accordingly if we do not normalize $\hat{\phi}(t)$ and $\mu \phi(t-t_{0})$ to 1 when considering their divergence in eq.~\eqref{eq:kl},
\begin{equation}
  \begin{aligned}
    D_\mathrm{KL}\left[\hat{\phi}(t) \parallel \mu\phi(t-t_0)\right] & =\int \left[\hat{\phi}(t) \log\frac{\hat{\phi}(t)}{\mu\phi(t-t_0)} + \mu\phi(t-t_0) - \hat{\phi}(t) \right]\mathrm{d}t \\
    & = - \int \hat{\phi}(t) \log\phi(t-t_0)\mathrm{d}t - \log(\mu)\int\hat{\phi}(t)\mathrm{d}t + \mu + \int \left[\hat{\phi}(t) \log\hat{\phi}(t) - \hat{\phi}(t) \right]\mathrm{d}t \\
    & = - \sum_{i=1}^{\hat{N}_\mathrm{PE}}\left[\int \hat{q}_i\delta(t-\hat{t_i}) \log\phi(t-t_0)\mathrm{d}t - \log(\mu)\int\hat{q}_i\delta(t-\hat{t_i})\mathrm{d}t\right] + \mu +  C \\
    & = -\log \left\{\prod_{i=1}^{\hat{N}_\mathrm{PE}} \left[\phi(\hat{t}_i-t_0)\right]^{\hat{q}_i}\right\} - \log(\mu)\sum_{i=1}^{\hat{N}_\mathrm{PE}} \hat{q}_i + \mu + C
  \label{eq:kl}
  \end{aligned}
\end{equation}
where $C$ is a constant regarding $t_0$ and $\mu$.  Define the time KL estimator as
\begin{equation}
  \begin{aligned}
  \label{eq:pseudo}
  \hat{t}_\mathrm{KL} &= \arg\underset{t_0}{\min}~D_\mathrm{KL}\left[\hat{\phi}(t) \parallel \mu\phi(t-t_0)\right] \\
  &= \arg\underset{t_0}{\max} \prod_{i=1}^{\hat{N}_\mathrm{PE}} \left[\phi(\hat{t}_i-t_0)\right]^{\hat{q}_i},
  \end{aligned}
\end{equation}
which reduces to an MLE like eq.~\eqref{eq:2} if $\hat{q}_i\equiv 1$.  $\hat{t}_\mathrm{KL}$ estimates $t_0$ when $t_i, q_i, N_\mathrm{PE}$ are all uncertain.
Similar to $\hat{t}_\mathrm{1st}$ and $\hat{t}_\mathrm{ALL}$, we define the standard deviation $\sqrt{\Var[\hat{t}_\mathrm{KL} - t_0]}$ to the resolution of an algorithm via KL divergence.

The intensity KL estimator is,
\begin{equation}
  \label{eq:pseudo-mu}
  \hat{\mu}_\mathrm{KL} = \arg\underset{\mu}{\min}~D_\mathrm{KL}\left[\hat{\phi}(t) \parallel \mu\phi(t-t_0)\right] = \sum_{i=1}^{\hat{N}_\mathrm{PE}} \hat{q}_i.
\end{equation}

\subsubsection{Residual sum of squares}
\label{sec:rss}

Following eqs.~\eqref{eq:1} and~\eqref{eq:lc}, we construct an estimator of a waveform,
\begin{equation}
  \label{eq:w-hat}
  \hat{w}(t) = \sum_{i=1}^{\hat{N}_\mathrm{PE}}\hat{q}_i V_\mathrm{PE}(t-\hat{t}_i) = \hat{\phi}(t) \otimes V_\mathrm{PE}(t).
\end{equation}

For a noise-free evaluation of $\hat{w}(t)$, residual sum of squares~(RSS) is a $\ell_2$-distance of it to $\tilde{w}(t)$,
\begin{equation}
  \label{eq:rss}
  \mathrm{RSS} \coloneqq\int\left[\hat{w}(t) - \tilde{w}(t)\right]^2\mathrm{d}t.
\end{equation}
We choose $\tilde{w}(t)$ for evaluating algorithms because otherwise with the raw waveform $w(t)$ RSS would be dominated by the white noise term $\epsilon(t)$.

\begin{figure}[H]
  \centering
  \resizebox{0.6\textwidth}{!}{\input{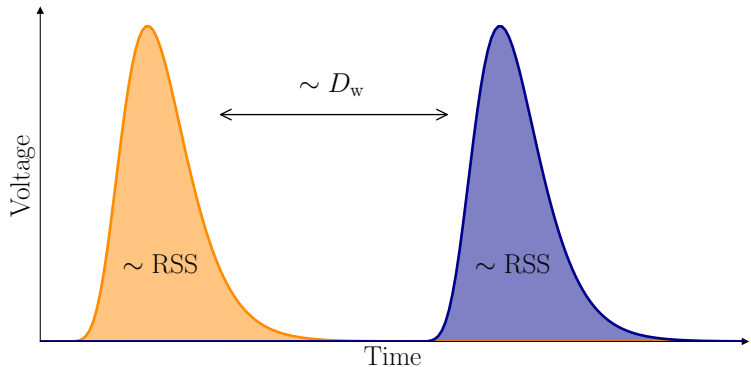}}
  \caption{\label{fig:l2} The $\mathrm{RSS}$ of orange and blue curves is a function of the two shaded regions. It is a constant when the curves shift horizontally when they do not overlap.  In contrast, the Wasserstein distance $D_\mathrm{w}$ of the two curves is associated with their separation.  It complements $\mathrm{RSS}$ and offers a time-sensitive metric suitable for the sparse PE space.}
\end{figure}

Figure~\ref{fig:l2} demonstrates that if two functions do not overlap, their $\mathrm{RSS}$ remain constant regardless of relative positions.  The delta functions in the sampled light curves $\hat{\phi}(t)$ and $\tilde{\phi}(t)$ hardly overlap, rendering $\mathrm{RSS}$ useless.  Furthermore, RSS cannot compare a discrete function with a continuous one.  We shall only consider the $\mathrm{RSS}$ of waveforms.

\subsubsection{Wasserstein distance}
\label{sec:W-dist}

Wasserstein distance $D_\mathrm{w}$ is a metric between two distributions, either of which can be discrete or continuous. It can capture the difference between a waveform analysis result $\hat{\phi}$ and the sampled light curve $\tilde{\phi}(t)$ in eq.~\eqref{eq:lc-sample}.
\begin{equation}
  D_\mathrm{w}\left[\hat{\phi}_*, \tilde{\phi}_*\right] = \inf_{\gamma \in \Gamma} \left[\int \left\vert t_1 - t_2 \right\vert \gamma(t_1, t_2)\mathrm{d}t_1\mathrm{d}t_2\right],
\end{equation}
where $*$ denotes the normalized light curves and $\Gamma$ is the collection of joint distributions with marginals $\hat{\phi}_*(t)$ and $\tilde{\phi}_*(t)$,
\begin{equation*}
  \label{eq:joint}
  \Gamma = \left\{\gamma(t_1, t_2) ~\middle\vert~ \int\gamma(t_1,t_2)\mathrm{d}t_1 = \tilde{\phi}_*(t_2) , \int\gamma(t_1,t_2)\mathrm{d}t_2 = \hat{\phi}_*(t_1) \right\}.
\end{equation*}
It is also known as the \textit{earth mover's distance}~\cite{levina_earth_2001}, encoding the minimum cost to transport mass from one distribution to another in figure~\ref{fig:l2}.

Alternatively, we can calculate $D_\mathrm{w}$ from cumulative distribution functions (CDF). Let $\hat\Phi(t)$ and $\tilde\Phi(t)$ denote the CDF of $\hat{\phi}_*(t)$ and $\tilde{\phi}_*(t)$, respectively. Then $D_\mathrm{w}$ is equivalent to the $\ell_1$-metric between the two CDFs,
\begin{equation}
    D_\mathrm{w}\left[\hat{\phi}_*, \tilde{\phi}_*\right] = \int\left|\hat{\Phi}(t) - \tilde{\Phi}(t)\right| \mathrm{d}t.
    \label{eq:numerical}
\end{equation}

In the following, we assess the performance of waveform analysis algorithms ranging from heuristics, deconvolution, neural network to regression by the criteria discussed in this section.

\subsection{Heuristic methods}
By directly extracting the patterns in the waveforms, \textit{heuristics} refer to the methods making minimal assumptions of the instrumental and statistical features.  Straightforward to implement and widely deployed in neutrino and dark matter experiments~\cite{students22}, they are poorly documented in the literature.  In this section, we try to formulate the heuristics actually have been used in the experiments so as to make an objective comparison with more advanced techniques.

\subsubsection{Waveform shifting}
\label{sec:shifting}
Some experiments use waveforms as direct input of analysis. Proton decay search at KamLAND~\cite{kamland_collaboration_search_2015} summed up all the PMT waveforms after shifting by time-of-flight for each event candidate.  The total waveform shape was used for a $\chi^2$-based particle identification (PID). The Double Chooz experiment also superposed waveforms to extract PID information by Fourier transformation~\cite{chooz_2018}. Samani~et~al.\cite{samani_pulse_2020} extracted pulse width from a raw waveform and use it as a PID discriminator.  Such techniques are extensions of pulse shape discrimination~(PSD) to large neutrino and dark matter experiments.  In the view of this study, extended PSD uses shifted waveform to approximate PE hit pattern, thus named \textit{waveform shifting}.

\begin{figure}[H]
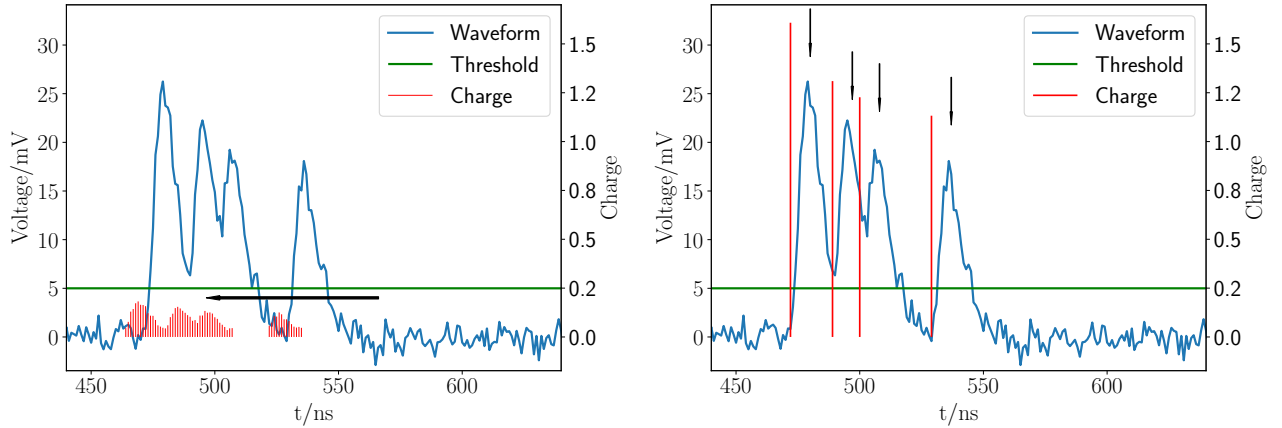

  \begin{subfigure}{.5\textwidth}
    \centering
    \resizebox{\textwidth}{!}{\input{figures/threshold.pgf}}
    \caption{\label{fig:shifting} A waveform shifting example gives \\ $\hat{t}_\mathrm{KL}-t_0=\SI{2.70}{ns}$, $\mathrm{RSS}=\SI{948.5}{mV^2}$, $D_\mathrm{w}=\SI{3.20}{ns}$.}
  \end{subfigure}
  \begin{subfigure}{.5\textwidth}
    \centering
    \resizebox{\textwidth}{!}{\input{figures/findpeak.pgf}}
    \caption{\label{fig:peak} A peak finding example gives \\ $\hat{t}_\mathrm{KL} - t_0=\SI{4.85}{ns}$, $\mathrm{RSS}=\SI{739.9}{mV^2}$, $D_\mathrm{w}=\SI{2.35}{ns}$.}
  \end{subfigure}
  \caption{\label{fig:method}Demonstrations of heuristic methods on a waveform sampled from $\mu=4$, $\tau_\ell=\SI{20}{ns}$, $\sigma_\ell=\SI{5}{ns}$ light curve conditions.  Peak finding in~\subref{fig:peak} handles charges more realistically than waveform shifting in~\subref{fig:shifting}, giving better numbers by the $\mathrm{RSS}$ and $D_\mathrm{w}$ criteria in section \ref{sec:criteria}. }
\end{figure}

As illustrated in figure~\ref{fig:shifting}, we firstly select all the $t_i$'s where the waveform $w(t_i)$ exceeds a threshold $V_\mathrm{th}$ to suppress noise, and shift them by a constant $\Delta t$. For an SER pulse $V_\mathrm{PE}(t)$ whose truth PE time is $t=0$, $\Delta t$ should minimize the Wasserstein distance $D_\mathrm{w}$. Thus,
\begin{equation}
    \Delta t \equiv \arg\underset{\Delta t'}{\min} D_\mathrm{w}\left[ V_\mathrm{PE*}(t), \delta(t-\Delta t') \right] \implies \int_{0}^{\Delta t} V_\mathrm{PE}(t) \mathrm{d}t = \frac{1}{2} \int_{0}^{\infty} V_\mathrm{PE}(t) \mathrm{d}t.
  \label{eq:waveform-shift-dt}
\end{equation}
The PE times are inferred as $\hat{t}_i = t_i - \Delta t$.  Corresponding $w(t_i)$'s are scaled by $\alpha$ to minimize RSS:
\begin{equation}
  \hat{\alpha} = \arg\underset{\alpha}{\min}~\mathrm{RSS}\left[ \alpha \sum_iw(t_i) \otimes V_\mathrm{PE}(t-\hat{t}_i), w(t) \right] .
  \label{eq:alpha}
\end{equation}
The charges are determined as $\hat{q}_i = \hat{\alpha} w(t_i)$.  Notice the difference from eq.~\eqref{eq:rss}: $\tilde{w}(t)$ unknown in data analysis, we replace it with $w(t)$.

Since the whole over-threshold waveform sample points are treated as PEs, one PE can be split into many. Thus, the obtained $\hat{q}_i$ are smaller than true PE charges. The waveform shifting model formulated above captures the logic behind waveform superposition methods.  The underlying assumption to treat a waveform as PEs is simply not true and time precision suffers.  It works only if the width of $V_\mathrm{PE}$ is negligible for the purpose, sometimes when classifying events.

\subsubsection{Peak finding}
\label{sec:findpeak}

The peak of $V_\mathrm{PE}$ is a distinct feature in waveforms, making \textit{peak finding} more effective than waveform shifting.  We smooth a waveform by a low-pass Savitzky-Golay filter~\cite{savitzky_smoothing_1964} and find all the peaks at $t_i$'s.  The following resembles waveform shifting: apply a constant shift $\Delta t \equiv \arg\underset{t}{\max} V_\mathrm{PE}(t)$ to get $\hat{t}_i = t_i - \Delta t$, and calculate a scaling factor $\alpha$ to get $\hat{q_i}=\hat{\alpha} w(t_i)$ in the same way as eq.~\eqref{eq:alpha}.  As shown in figure~\ref{fig:peak}, peak finding outputs charges close to 1 and works well for lower PE counts.  But when PEs pile up closely, peaks overlap intensively, making this method unreliable.  Peak finding is usually too trivial to be documented but found almost everywhere~\cite{students22}.

\subsection{Deconvolution}
\label{sec:deconv}
Deconvolution is motivated by viewing the waveform as a convolution of sparse spike train $\tilde{\phi}$ and $V_\mathrm{PE}$ in eq.~\eqref{eq:1}.  Huang et al.~\cite{huang_flash_2018} from DayaBay and Grassi et al.~\cite{grassi_charge_2018} introduced deconvolution-based waveform analysis in charge reconstruction and linearity studies.  Zhang et al.~\cite{zhang_comparison_2019} then applied it to the JUNO prototype.  Deconvolution methods are better than heuristic ones by using the full shape of $V_\mathrm{PE}(t)$, thus can accommodate overshoots and pile-ups.  Noise and Nyquist limit make deconvolution sensitive to fluctuations in real-world applications.  A carefully selected low-pass filter mitigates the difficulty but might introduce Gibbs ringing artifacts in the smoothed waveforms and the deconvoluted results. Despite such drawbacks, deconvolution algorithms are fast and useful to give initial crude solutions for the more advanced algorithms.  Deployed in running experiments, they are discussed in this section to make an objective evaluation. 

\subsubsection{Fourier deconvolution}
\label{sec:fourier}
The deconvolution relation is evident in the Fourier transform $\mathcal{F}$ to eq.~\eqref{eq:1},
\begin{equation}
  \label{eq:fourier}
  \mathcal{F}[w]  = \mathcal{F}[\tilde{\phi}]\mathcal{F}[V_\mathrm{PE}] + \mathcal{F}[\epsilon]
  \implies \mathcal{F}[\tilde{\phi}]  = \frac{\mathcal{F}[w]}{\mathcal{F}[V_\mathrm{PE}]} - \frac{\mathcal{F}[\epsilon]}{\mathcal{F}[V_\mathrm{PE}]}.
\end{equation}
By low-pass filtering the waveform $w(t)$ to get $\tilde{w}(t)$, we suppress the noise term $\epsilon$.  In the inverse Fourier transform $\hat{\phi}_1(t) = \mathcal{F}^{-1}\left[\frac{\mathcal{F}[\tilde{w}]}{\mathcal{F}[V_\mathrm{PE}]}\right](t)$, remaining noise and limited bandwidth lead to smaller and even negative $\hat{q}_i$.  We apply a $q_\mathrm{th}$ threshold regularizer to cut off the unphysical parts of $\hat{\phi}_1(t)$,
\begin{equation}
  \label{eq:fdconv2}
    \hat{\phi}(t) = \hat{\alpha}\underbrace{\hat{\phi}_1(t) I\left(\hat{\phi}_1(t) - q_\mathrm{th}\right)}_{\text{over-threshold part of} \hat{\phi}_1(t)}  
\end{equation}
where $I(x)$ is the indicator function, and $\hat{\alpha}$ is the scaling factor to minimize $\mathrm{RSS}$ like in eq.~\eqref{eq:alpha},
\begin{equation*}
  \begin{aligned}
  \label{eq:id}
  I(x) = \left\{
    \begin{array}{ll}
      1 & \mbox{, if $x\ge0$}, \\
      0 & \mbox{, otherwise}
    \end{array}
    \right.
    \quad~~~
    \hat{\alpha} = \arg \underset{\alpha}{\min}\mathrm{RSS}\left[\alpha \hat{\phi} \otimes V_\mathrm{PE}, w\right]. \\
  \end{aligned}
\end{equation*}

\begin{figure}[H]
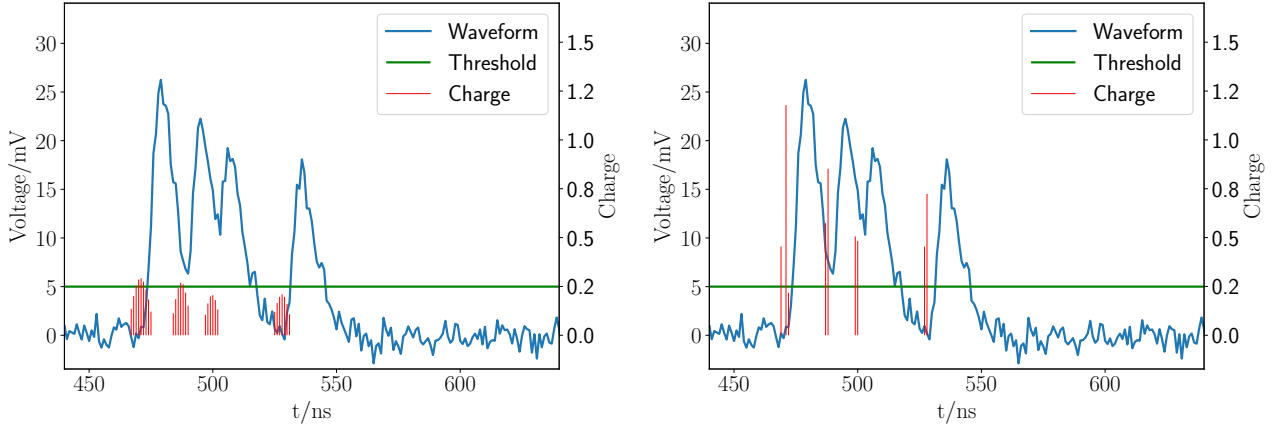

  \begin{subfigure}{0.5\textwidth}
    \centering
    \resizebox{\textwidth}{!}{\input{figures/fftrans.pgf}}
    \caption{\label{fig:fd} A Fourier deconvolution example: \\ $\hat{t}_\mathrm{KL} - t_0=\SI{2.61}{ns}$, $\mathrm{RSS}=\SI{153.7}{mV^2}$, $D_\mathrm{w}=\SI{1.87}{ns}$.}
  \end{subfigure}
  \begin{subfigure}{0.5\textwidth}
    \centering
    \resizebox{\textwidth}{!}{\input{figures/lucyddm.pgf}}
    \caption{\label{fig:lucy} A Richardson-Lucy direct demodulation example:\\ $\hat{t}_\mathrm{KL} - t_0=\SI{2.77}{ns}$, $\mathrm{RSS}=\SI{10.0}{mV^2}$, $D_\mathrm{w}=\SI{0.60}{ns}$.}
  \end{subfigure}
  \caption{\label{fig:deconv}Demonstrations of deconvolution methods on a waveform sampled from the same setup as figure~\ref{fig:method}. Richardson-Lucy direct demodulation in~\subref{fig:lucy} imposes positive charges in iterations and obtains better results than Fourier deconvolution in~\subref{fig:fd}.}
\end{figure}

Figure~\ref{fig:fd} illustrates that Fourier deconvolution outperforms heuristic methods, but still with a lot of small-charged PEs.

\subsubsection{Richardson-Lucy direct demodulation}
\label{sec:lucyddm}

\textit{Richardson-Lucy direct demodulation}~(LucyDDM)~\cite{lucy_iterative_1974} with a non-linear iteration to calculate deconvolution has a wide application in astronomy~\cite{li_richardson-lucy_2019} and image processing. We view $V_{\mathrm{PE}*}(t-s)$ as a conditional probability distribution $p(t|s)$ where $t$ denotes PMT amplified electron time, and $s$ represents the given PE time. By the Bayesian rule,
\begin{equation}
  \label{eq:lucy}
  \tilde{\phi}_*(s) V_{\mathrm{PE}*}(t-s) = \tilde{\phi}_*(s)p(t|s) = p(t,s) = \tilde{w}_*(t)p(s|t),
\end{equation}
where $p(t, s)$ is the joint distribution of amplified electron $t$ and PE time $s$, and $\tilde{w}$ is the smoothed $w$.  Cancel out the normalization factors,
\begin{equation}
  \label{eq:ptt}
  p(s|t) = \frac{\tilde{\phi}_*(s) V_{\mathrm{PE}*}(t-s)}{\tilde{w}_*(t)} = \frac{\tilde{\phi}(s) V_{\mathrm{PE}}(t-s)}{\int\tilde{\phi}(s') V_{\mathrm{PE}}(t-s')\mathrm{d}s'}.
\end{equation}
Then a recurrence relation for $\phi_*$ is,
\begin{equation}
  \label{eq:iter}
  \begin{aligned}
    \tilde{\phi}_*(s) & = \int p(s|t) \tilde{w}_*(t)\mathrm{d}t = \int \frac{\tilde{\phi}(s) V_{\mathrm{PE}}(t-s)}{\int\tilde{\phi}(s') V_{\mathrm{PE}}(t-s')\mathrm{d}s'} \tilde{w}_*(t) \mathrm{d}t \\
    \implies \hat{\phi}^{n+1}(s) & = \int \frac{\hat{\phi}^n(s) V_{\mathrm{PE}*}(t-s)}{\int\hat{\phi}^n(s') V_{\mathrm{PE}}(t-s')\mathrm{d}s'} \tilde{w}(t) \mathrm{d}t,
  \end{aligned}
\end{equation}
where only $V_{\mathrm{PE}*}$ in the numerator is normalized, and superscript $n$ denotes the iteration step.
Like Fourier deconvolution in eq.~\eqref{eq:fdconv2}, we threshold and scale the converged $\hat{\phi}^\infty$ to get $\hat{\phi}$.  As shown in figure~\ref{fig:lucy}, the positive constraint of $\hat{\phi}$ makes LucyDDM more resilient to noise.

The remaining noise in the smoothed $\tilde{w}$ crucially influences deconvolution.  A probabilistic method will correctly model the noise term $\epsilon$, as we shall see in section \ref{sec:regression}.

\subsection{Convolutional neural network}
\label{sec:cnn}
Convolutional neural networks~(CNN) made breakthroughs in various fields like computer vision~\cite{he_deep_2016} and natural language processing~\cite{vaswani_attention_2017}.  As an efficient composition of weighted additions and non-linear functions, neural networks outperform many traditional algorithms.  The success of CNN induces many ongoing efforts to apply it to waveform analysis~\cite{students22}.  It is thus interesting and insightful to make a comparison of CNN with the remaining traditional methods.

The input discretized waveform $\bm{w}$ is a 1-dimensional vector.  However, $\hat{q}_i$ and $\hat{t}_i$ are two variable length ($\hat{N}_\mathrm{PE}$) vectors, which is not a well-defined output for a 1-dimensional CNN~(1D-CNN).  Instead, we replace $\hat{N}_\mathrm{PE}$ with a fixed sample size $N_\mathrm{s}$ and $\hat{t}_i$ with a fixed grid of times $t'_j$ associating $q'_j$. For most $j$, $q'_j = 0$, meaning there is no PE on time grid $t'_j$. By stripping out $j$ where $q'_j=0$, the remaining $q'_j$, $t'_j$ are $\hat{q}_i$ and $\hat{t}_i$.  Now $q'_j$ is a 1D vector with fixed length $N_\mathrm{s}$, suitable for 1D-CNN.

\begin{figure}[H]
  \begin{subfigure}{.4\textwidth}
    \centering
    \begin{adjustbox}{width=0.5\textwidth}
      \tikzstyle{block} = [rectangle, rounded corners, minimum width=2cm, minimum height=1cm, text centered, draw=black]
\tikzstyle{arrow} = [thick, ->, >=stealth]
\begin{tikzpicture}[node distance=2cm]
    \node (0) [block] {$1\times1029$};
    \node (1) [block, below of=0] {$25\times1029$};
    \node (2) [block, below of=1] {$20\times1029$};
    \node (3) [block, below of=2] {$15\times1029$};
    \node (4) [block, below of=3] {$10\times1029$};
    \node (5) [block, below of=4] {$1\times1029$};
    \draw [arrow] (0) -- node [midway](0to1) {} (1);
    \draw [arrow] (1) -- node [midway](1to2) {} (2);
    \draw [arrow] (2) -- node [midway](2to3) {} (3);
    \draw [arrow] (3) -- node [midway](3to4) {} (4);
    \draw [arrow] (4) -- node [midway](4to5) {} (5);
    \node (a) [block, right of=0to1, xshift=1.2cm] {kernel=$21$};
    \node (b) [block, right of=1to2, xshift=1.2cm] {kernel=$17$};
    \node (c) [block, right of=2to3, xshift=1.2cm] {kernel=$13$};
    \node (d) [block, right of=3to4, xshift=1.2cm] {kernel=$9$};
    \node (e) [block, right of=4to5, xshift=1.2cm] {kernel=$1$};
    \draw [arrow] (a) -- (0to1) {};
    \draw [arrow] (b) -- (1to2) {};
    \draw [arrow] (c) -- (2to3) {};
    \draw [arrow] (d) -- (3to4) {};
    \draw [arrow] (e) -- (4to5) {};
\end{tikzpicture}
    \end{adjustbox}
    \caption{\label{fig:struct} Structure of the neural network.}
  \end{subfigure}
  \begin{subfigure}{.5\textwidth}
    \centering
    \resizebox{\textwidth}{!}{\input{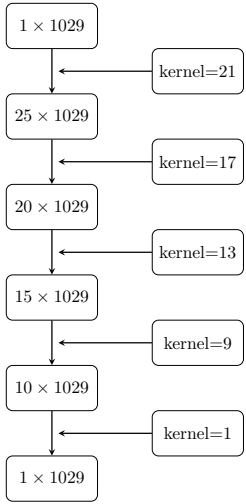}
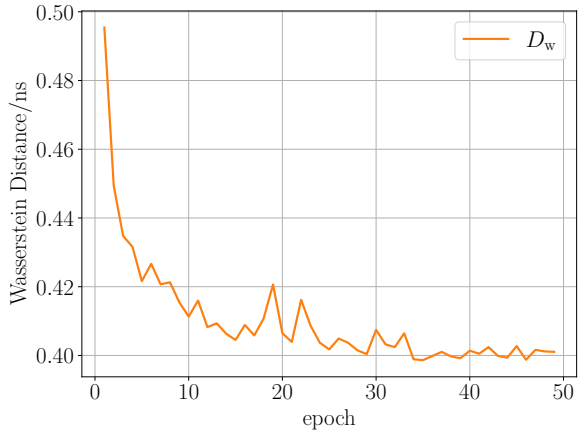}
    \caption{\label{fig:loss} Evolution of loss.}
  \end{subfigure}
  \caption{\label{fig:CNN} Training process of a CNN. A shallow network structure of 5 layers in~\subref{fig:struct} is trained to converge in Wasserstein distance as shown in~\subref{fig:loss}.  ``kernel=21'' stands for a 1-D convolutional kernel size 21. ``1029'' is the number of voltage samples in a waveform.  $1\times$ represents the number of channels in each layer.}
\end{figure}

We choose a shallow network structure of 5 layers to recognize patterns as shown in figure~\ref{fig:struct}, motivated by the pulse shape and universality of $V_\mathrm{PE}(t)$ for all the PEs. The input waveform vector $\bm{w}$ is convolved by several kernels $\bm{K}^{(1)}_m$ into new vectors $v_m$:
\begin{equation}
  \bm{v}^{(1)}_m = \bm{K}^{(1)}_m \otimes \bm{w},\ m\in \{1,\ldots,M\}.
  \label{eq:1DCNN-11}
\end{equation}
As 1D vectors, $\bm{K}^{(1)}_m$ share the same length called \textit{kernel size}.  $M$ is the \textit{number of channels}. As shown in figure~\ref{fig:struct}, considering the localized nature of $V_\mathrm{PE}(t)$,  we choose the kernel size to be $21$ and $M=25$.

After the above linear \textit{link} operations, a point-wise nonlinear \textit{activation} transformation, leaky rectified linear function $\mathrm{LReL}(\cdot)$\cite{leakyReLU} is used:
\begin{equation}
  \begin{aligned}
    & \bm{v'}^{(1)}_m = \mathrm{LReL}(\bm{v}^{(1)}_m) \\
    \text{where  } & \mathrm{LReL}(x) = \left\{ \begin{aligned}
      & 0.05 x & x<0 \\
      & x & x\geqslant 0 \\
    \end{aligned} \right.
  \end{aligned}
  \label{eq:1DCNN-12}
\end{equation}
The two operations form the first layer. The second layer is similar,
\begin{equation}
  \bm{v}'^{(2)}_n = \mathrm{LReL}\left(\sum_{m=1}^{M} \bm{K}^{(2)}_{nm} \otimes \bm{v'}^{(1)}_m\right),\ n\in \{1,\ldots,N\},
  \label{eq:1DCNN-2}
\end{equation}
mapping $M$-channeled $\bm{v'}^{(1)}_m$ to $N$-channeled $\bm{v'}^{(2)}_n$.

At the bottom of figure~\ref{fig:struct}, 1D-CNN gives the desired output, a one-channeled vector $q'_j$, which determines the PE distribution $\phi'(t)$ by
\begin{equation}
  \label{eq:gd-phi}
  \phi'(t) = \sum_{j=1}^{N_\mathrm{s}}q'_j\delta(t-t'_j).
\end{equation}

The whole network is a non-linear function $\mathrm{CNN}(\cdot)$ from $\bm{w}$ to $\phi'$ with numerous free parameters $\bm{K}^{(1)}_m, \bm{K}^{(2)}_{mn}, \ldots$ which we denote as $\mathcal{K}$. We \textit{train} to fit the parameters against true $\tilde{\phi}$,
\begin{equation}
  \hat{\mathcal{K}} = \arg\underset{\mathcal{K}}{\min} D_\mathrm{w}\left[\mathrm{CNN}(\bm{w}; \mathcal{K}), \tilde{\phi}\right]
  \label{eq:CNN-train}
\end{equation}
by back-propagation. Figure~\ref{fig:loss} shows the convergence of Wasserstein distance during training. Such fitting process is an example of \textit{supervised learning}. As explained in figure~\ref{fig:l2}, $D_\mathrm{w}$ can naturally measure the time difference between two sparse $\phi'$ and $\tilde{\phi}$ in eq.~\eqref{eq:CNN-train}, making 1D-CNN not need to split a PE into smaller ones to fit waveform fluctuations.  This gives sparser results in contrast to deconvolution methods in section~\ref{sec:deconv} and direct charge fitting in section~\ref{sec:dcf}, which shall be further  discussed in section~\ref{sec:sparsity}.

\begin{figure}[H]
  \begin{subfigure}{.5\textwidth}
    \centering
    \resizebox{\textwidth}{!}{\input{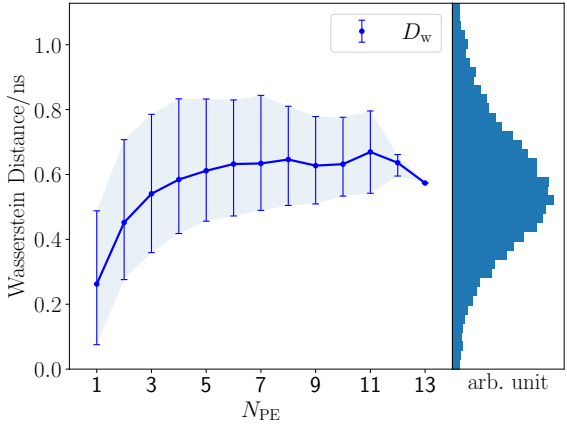}}
    \caption{\label{fig:cnn-npe} $D_\mathrm{w}$ histogram and its distributions conditioned \\ on $N_{\mathrm{PE}}$. ``arbi. unit'' means arbitrary unit.}
  \end{subfigure}
  \begin{subfigure}{.5\textwidth}
    \centering
    \resizebox{\textwidth}{!}{\input{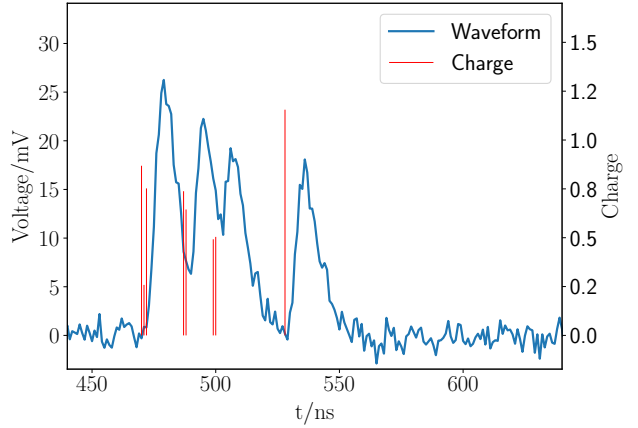}}
    \caption{\label{fig:cnn}An example giving \\ $\hat{t}_\mathrm{KL} - t_0=\SI{2.96}{ns}$, $\mathrm{RSS}=\SI{19.5}{mV^2}$, $D_\mathrm{w}=\SI{0.81}{ns}$.}
  \end{subfigure}
  \caption{\label{fig:cnn-performance}Demonstration of CNN on $\num[retain-unity-mantissa=false]{1e4}$ waveforms in~\subref{fig:cnn-npe} and one waveform in~\subref{fig:cnn} sampled from the same setup as figure~\ref{fig:method}. In figure~\subref{fig:cnn-npe}, the middle line is the mean of the distribution. The size of errorbar is from \SIrange{15.8}{84.1}{\percent} quantiles, corresponding to $\SI{\pm 1}{\sigma}$ of a Gaussian distribution. }
\end{figure}

In figure~\ref{fig:cnn-npe}, $D_\mathrm{w}$ is the smallest for one PE.  $D_\mathrm{w}$ stops increasing with $N_\mathrm{PE}$ at about 6 PEs.  When $N_\mathrm{PE}$ is more than 6, pile-ups tend to produce a continuous waveform and the average PE time accuracy stays flat. Similar to eq.~\eqref{eq:fdconv2}, the output of CNN should be scaled by $\hat{\alpha}$ to get $\hat{\phi}$. Such small $D_\mathrm{w}$ in figure~\ref{fig:cnn-npe} provides a precise matching of waveforms horizontally in the time axis to guarantee effective $\hat{\alpha}$ scaling, explaining why $\mathrm{RSS}$ is also small in figure~\ref{fig:cnn}.

\subsection{Regression analysis}
\label{sec:regression}
With the generative waveform model in eq.~\eqref{eq:1}, regression is ideal for analysis. Although computational complexity hinders the applications of regression by the vast volumes of raw data, the advancement of sparse models and big data infrastructures strengthens the advantage of regression.

The truth $N_\mathrm{PE}$ is unknown and formulating an explicit trans-dimensional model is expansive.  So, in the first two methods, we use the grid representation of PE sequence $q'_j, j\in \{1\cdots N_\mathrm{s}\}$ introduced in \ref{sec:cnn} in order to avoid cross-dimensional issues.  We shall solve the issue and turn back to length-varying representation in section~\ref{subsec:fsmp}.

Regression methods adjust $\{q'_j\}$ to fit eq.~\eqref{eq:gd}:
\begin{equation}
  \label{eq:gd}
  w'(t) = \sum_{j=1}^{N_\mathrm{s}}q'_jV_\mathrm{PE}(t-t'_j).
\end{equation}

From the output $\hat{\phi}_\mathrm{dec}(t)$ of a deconvolution method in section~\ref{sec:lucyddm}, we confidently leave out all the $t'_j$ that $\hat{\phi}_\mathrm{dec}(t_j')=0$ in eq.~\eqref{eq:gd-phi} to reduce the number of parameters and the complexity.

\subsubsection{Direct charge fitting}
\label{sec:dcf}

Fitting the charges $q'_j$ in eq.~\eqref{eq:gd} directly by minimizing RSS of $w'(t)$ and $w(t)$, we get
\begin{equation}
  \label{eq:gd-q}
  \bm{\hat{q}} = \arg \underset{q'_j \ge 0}{\min}~\mathrm{RSS}\left[w'(t),w(t)\right].
\end{equation}
RSS of eq.~\eqref{eq:gd-q} does not suffer from the sparse configuration in figure~\ref{fig:l2} provided that the dense grid in eq.~\eqref{eq:gd} covers all the PEs.

\begin{figure}[H]
  \begin{subfigure}{.5\textwidth}
    \centering
    \resizebox{\textwidth}{!}{\input{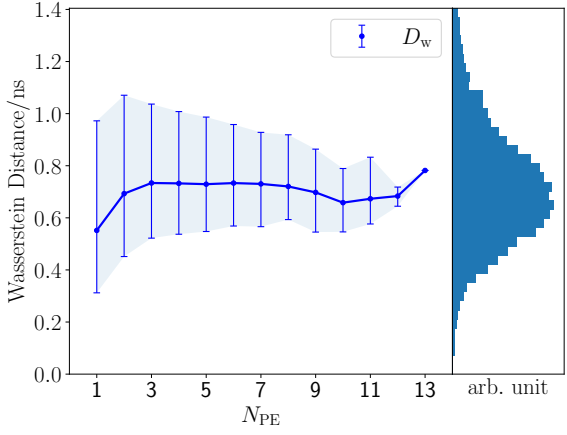}}
    \caption{\label{fig:fitting-npe} $D_\mathrm{w}$ histogram and its distributions conditioned \\ on $N_{\mathrm{PE}}$, errorbar explained in figure~\ref{fig:cnn-performance}.}
  \end{subfigure}
  \begin{subfigure}{.5\textwidth}
    \centering
    \resizebox{\textwidth}{!}{\input{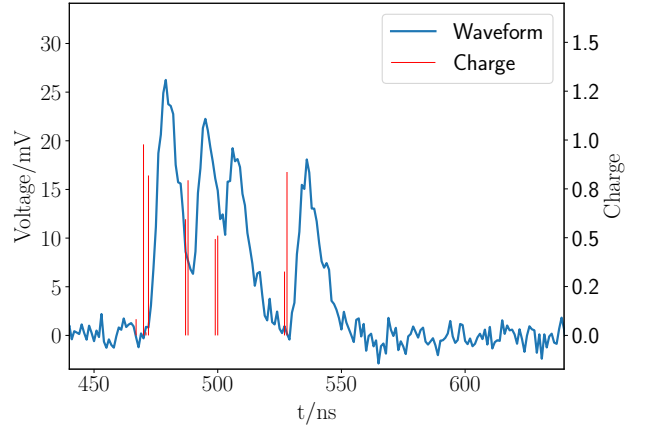}}
    \caption{\label{fig:fitting}An example giving \\ $\hat{t}_\mathrm{KL} - t_0=\SI{2.73}{ns}$, $\mathrm{RSS}=\SI{6.49}{mV^2}$,$D_\mathrm{w}=\SI{0.91}{ns}$.}
  \end{subfigure}
  \caption{\label{fig:dcf}Demonstration of direct charge fitting with $\num[retain-unity-mantissa=false]{1e4}$ waveforms in~\subref{fig:fitting-npe} and one waveform in~\subref{fig:fitting} sampled from the same setup as figure~\ref{fig:method}.  Direct charge fitting shows a better performance than LucyDDM in figure~\ref{fig:lucy} and a comparable $D_\mathrm{w}$ to CNN in figure~\ref{fig:cnn}.}
\end{figure}

Slawski and Hein~\cite{slawski_non-negative_2013} proved that in deconvolution problems, the non-negative least-squares formulation in eq.~\eqref{eq:gd-q} is self-regularized and gives sparse solutions of $q'_i$.  Peterson~\cite{peterson_developments_2021} from IceCube used this technique for waveform analysis.  We optimize eq.~\eqref{eq:gd-q} by Broyden-Fletcher-\allowbreak{}Goldfarb-Shanno algorithm with a bound constraint~\cite{byrd_limited_1995}.  In figure~\ref{fig:fitting-npe}, charge fitting is consistent in $D_\mathrm{w}$ for different $N_\mathrm{PE}$'s, showing its resilience to pile-up.

The sparsity of $q'_i$ is evident in figure~\ref{fig:fitting}.  However, the majority of the $\hat{q}_i$ are less than 1.  This feature motivates us to incorporate prior knowledge of $q'_i$ towards a more dedicated model than directly fitting charges.

\subsubsection{Hamiltonian Monte Carlo}
\label{subsec:mcmc}
Chaining the $q'_i$ distribution~(section~\ref{subsec:spe}), the charge fitting eq.~\eqref{eq:gd-q} and the light curve eq.~\eqref{eq:time-pro}, we arrive at a hierarchical Bayesian model,
\begin{equation}
  \begin{aligned}
    t_{0} &\sim \mathcal{U}(0, \overline{t_0}) \\
    \mu_j &= \mu \int_{t'_j-\frac{\Delta t'}{2}}^{t'_j+\frac{\Delta t'}{2}} \phi(t' - t_0)\mathrm{d}t' \approx \mu\phi(t'_j - t_0)\Delta{t'} \\
    z_j &\sim \mathcal{B}(\mu_j) \\
    q'_{j,0}&=0\\
    q'_{j,1}& \sim \Gamma(k=1/0.4^2, \theta=0.4^2)\\
    q'_j &= q'_{j,z_j}\\
    w'(t) & = \sum_{j=1}^{N_\mathrm{s}}q'_jV_\mathrm{PE}(t-t'_j)\\
    w(t) &\sim \mathcal{N}(w'(t), \Var[\epsilon])
  \end{aligned}
  \label{eq:mixnormal}
\end{equation}
where $\mathcal{U}$, $\mathcal{B}$ and $\Gamma$ stand for uniform, Bernoulli and gamma distributions, $\overline{t_0}$ is an upper bound of $t_0$, and $q'_j$ is a mixture of 0 (no PE) and gamma-distributed $q'_{j,1}$ (1 PE). When the expectation $\mu_j$ of a PE hitting $(t'_{j} - \frac{\Delta t'}{2}, t'_{j} + \frac{\Delta t'}{2})$ is small enough, that 0-1 approximation is valid.  The inferred waveform $w'(t)$ differs from observable $w(t)$ by a white noise $\epsilon(t) \sim \mathcal{N}(0, \Var[\epsilon])$ motivated by eq.~\eqref{eq:1}.  When an indicator $z_j=0$, it turns off $q'_j$, reducing the number of parameters by one.  That is how eq.~\eqref{eq:mixnormal} achieves sparsity.

We generate posterior samples of $t_0$ and $\bm{q'}$ by Hamiltonian Monte Carlo~(HMC)~\cite{neal_mcmc_2012}, a variant of Markov chain Monte Carlo suitable for high-dimensional problems. Construct $\hat{t}$ and $\hat{q}_j$ as the mean estimators of posterior samples $t_0$ and $q'_j$ at $z_j=1$.  Unlike the $\hat{t}_\mathrm{KL}$ discussed in section~\ref{sec:pseudo}, $\hat{t}_0$ is a direct Bayesian estimator from eq.~\eqref{eq:mixnormal}.  We construct $\hat{\phi}(t)$ by eq.~\eqref{eq:gd-phi} and $\hat{w}(t)$ by $\hat{\phi} \otimes V_\mathrm{PE}$. RSS and $D_\mathrm{w}$ are then calculated according to eqs.~\eqref{eq:rss} and \eqref{eq:numerical}.

\begin{figure}[H]
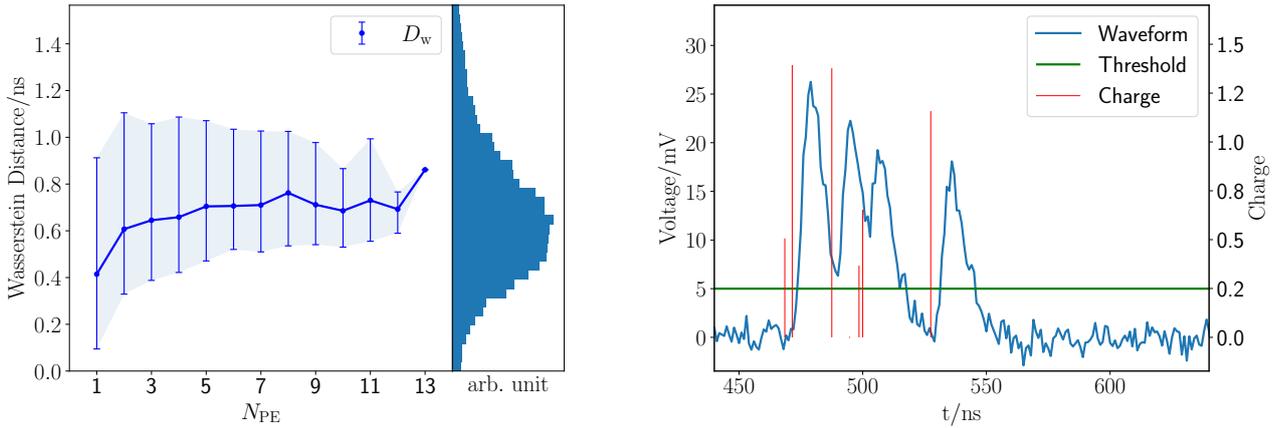

  \begin{subfigure}{.5\textwidth}
    \centering
    \resizebox{\textwidth}{!}{\input{figures/mcmcchargestats.pgf}}
    \caption{\label{fig:mcmc-npe} $D_\mathrm{w}$ histogram and its distributions conditioned \\ on $N_{\mathrm{PE}}$, errorbar explained in figure~\ref{fig:cnn-performance}.}
  \end{subfigure}
  \begin{subfigure}{.5\textwidth}
    \centering
    \resizebox{\textwidth}{!}{\input{figures/mcmc.pgf}}
    \caption{\label{fig:mcmc}An example with \\ $\hat{t}_0 - t_0=\SI{2.67}{ns}$, $\mathrm{RSS}=\SI{8.83}{mV^2}$, $D_\mathrm{w}=\SI{0.62}{ns}$.}
  \end{subfigure}
  \caption{\label{fig:mcmc-performance}Demonstration of HMC with $\num[retain-unity-mantissa=false]{1e4}$ waveforms in~\subref{fig:mcmc-npe} and one waveform in~\subref{fig:mcmc} sampled from the same setup as figure~\ref{fig:method}.  Although using a more dedicated model, HMC performs worse than the direct charge fitting in figure~\ref{fig:dcf}. We suspect the Markov chain is not long enough.}
\end{figure}
Although we imposed a prior distribution in eq.~\eqref{eq:mixnormal} with $\E[q_j]=1$, the charges $\hat{q}_j$ in figure~\ref{fig:mcmc} are still less than 1.  The $D_\mathrm{w}$ marginal distribution in figure~\ref{fig:mcmc-npe} is less smooth than that of the direct charge fitting in figure~\ref{fig:fitting-npe}.  Similarly, RSS in figure~\ref{fig:mcmc} is slightly worse than that in figure~\ref{fig:fitting}.  We suspect the Markov chain has not finally converged due to the trans-dimensional property of eq.~\eqref{eq:mixnormal}.  Extending the chain is not a solution because HMC is already much slower than direct fitting in section~\ref{sec:dcf}.  We need an algorithm that pertains to the model of eq.~\eqref{eq:mixnormal} but much faster than HMC.

\subsubsection{Fast stochastic matching pursuit}
\label{subsec:fsmp}
In reality, $w(t)$ is discretized as $\bm{w}$. If we rewrite the hierarchical model in eq.~\eqref{eq:mixnormal} into a joint distribution, marginalizing out $\bm{q}'$ and $\bm{z}$ gives a flattened model,
\begin{equation}
  \label{eq:universe}
  \begin{aligned}
    p(\bm{w}, \mu) &= \sum_{\bm{z}, t_0} \int \mathrm{d}\bm{q}' p(\bm{w}, \bm{q}', \bm{z}, t_0, \mu) \\
    &= p(\mu) \sum_{\bm{z}, t_0} \left[\int \mathrm{d}\bm{q}' p(\bm{w}|\bm{q}') p(\bm{q}'|\bm{z}, t_0) \right] p(\bm{z}, t_0|\mu) \\
    &= p(\mu) \sum_{\bm{z}, t_0} p(\bm{w}|\bm{z}, t_0) p(\bm{z}, t_0|\mu) \\
    &= p(\mu) p(\bm{w}|\mu) \\
  \end{aligned}
\end{equation}
The integration over $\bm{q}'$ is the probability density of a multi-normal distribution $p(\bm{w}|\bm{z}, t_0)$, with a fast algorithm to iteratively compute by Schniter~et al.~\cite{schniter_fast_2008}. The summation over $\bm{z}$ and $t_0$, however, takes an exploding number of combinations.

Let's approximate the summation with a sample from $S = \{(\bm{z}^1, t_{0}^1), (\bm{z}^2, t_{0}^2), \cdots, (\bm{z}^M, t_{0}^M)\}$ by Metropolis-Hastings~\cite{metropolis_equation_1953, hastings_monte_1970, mackay_information_2003} based Gibbs hybrid MCMC~\cite{tierney_1994}. A $t_{0}^i$ is sampled from $p(t_0|\bm{z}^{i-1}) \propto p(\bm{z}^{i-1}|t_0)p(t_0)$; a $\bm{z}^i$ is sampled from $p(\bm{z}|t_{0}^{i}) = C p(\bm{w} | \bm{z}, t_0) h(\bm{z}, t_0^i)$. $C$ is independent of $\bm{z}$, and $h(\bm{z}, t_0) = p(\bm{z}|\mu_0, t_0)p(t_0)$, where $\mu_0$ is an educated guess from a previous method like LucyDDM~(section~\ref{sec:lucyddm}), and $p(t_0)$ is the prior distribution of $t_0$. Then,
\begin{equation}
  \label{eq:mh}
  \begin{aligned}
    p(\bm{w}|\mu) &= \sum_{\bm{z}, t_0} p(\bm{w}|\bm{z}, t_0) p(\bm{z}, t_0|\mu) = \frac{1}{C}\sum_{\bm{z}, t_0} p(\bm{z}|t_0) \frac{p(\bm{z}, t_0 | \mu)}{h(\bm{z}, t_0)} \\
    &= \frac{1}{C} \E_{\bm{z}, t_0}\left[ \frac{p(\bm{z}, t_0 | \mu)}{h(\bm{z}, t_0)} \right] = \frac{1}{C} \E_{\bm{z},t_0}\left[ \frac{p(\bm{z} | \mu, t_0)p(t_0)}{p(\bm{z} | \mu_0, t_0)p(t_0)} \right]\\
    &\approx \frac{1}{CM} \sum_{i=1}^M \frac{p(\bm{z}^i | \mu, t_{0}^i)}{p(\bm{z}^i | \mu_0, t_{0}^i)}
  \end{aligned}
\end{equation}
Construct the approximate MLEs for $\mu$ and $\bm{z}$, and the expectation estimator of $t_0$ and $\hat{\bm{q}}$,
\begin{equation}
  \label{eq:fsmpcharge}
  \begin{aligned}
    \hat{t}_0 &= \frac{1}{M}\sum_{i=0}^M t_{0}^i\\
    \hat{\mu} &= \arg\underset{\mu}{\max}~p(\bm{w}|\mu) = \arg\underset{\mu}{\max} \sum_{i=1}^M \frac{p(\bm{z}^i | \mu, t_{0}^i)}{p(\bm{z}^i | \mu_0, t_{0}^i)}\\
    \hat{\bm{z}} &= \underset{(\bm{z}^i, t_{0}^i) \in S}{\arg\max}~p(\bm{w}|\bm{z}^i, t_0^i) h(\bm{z}^i, t_{0}^i) \\
    \hat{\bm{q}}|{\hat{\bm{z}}} &= \E(\bm{q}'|\bm{w},\hat{\bm{z}})
  \end{aligned}
\end{equation}
RSS and $D_\mathrm{w}$ are calculated by eqs.~\eqref{eq:rss}, \eqref{eq:numerical}, \eqref{eq:gd-phi}.

We name the method \emph{fast stochastic matching pursuit}~(FSMP) after \emph{fast Bayesian matching pursuit}~(FBMP) by Schniter~et al.~\cite{schniter_fast_2008} and \emph{Bayesian stochastic matching pursuit} by Chen~et~al.~\cite{chen_stochastic_2011}.  Here FSMP replaces the greedy search routine in FBMP with stochastic sampling.  With the help of Ekanadham~et al.'s function interpolation~\cite{ekanadham_recovery_2011}, FSMP straightforwardly extends $\bm{z}$ into an unbinned vector of PE locations $t_i$.  Geyer and Møller~\cite{geyer_simulation_1994} developed a similar sampler to handle trans-dimensionality in a Poisson point process.  $h(\bm{z}, t_0)$ and the proposal distribution in Metropolis-Hastings steps could be tuned to improve sampling efficiency.  We shall leave the detailed study of the Markov chain convergence to our future publications.

\begin{figure}[h]
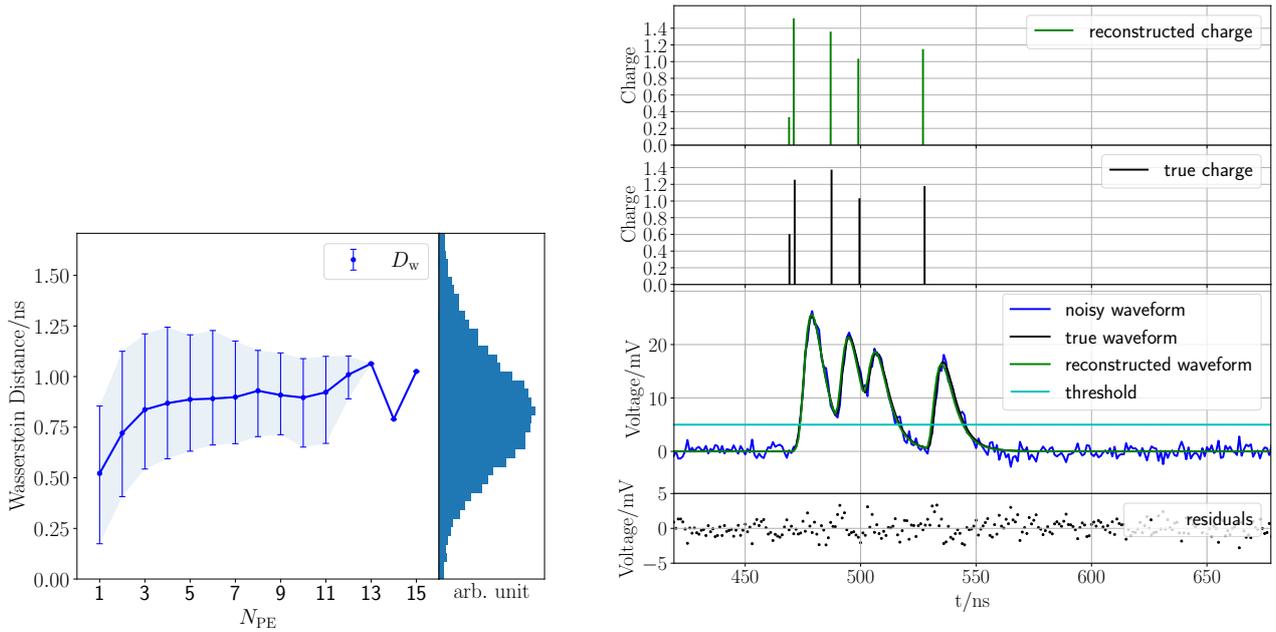

  \begin{subfigure}[b]{.45\textwidth}
    \centering
    \resizebox{1.05\textwidth}{!}{\input{figures/fsmpchargestats.pgf}}
    \caption{\label{fig:fsmp-npe} $D_\mathrm{w}$ histogram and distributions conditioned on $N_{\mathrm{PE}}$, errorbar explained in figure~\ref{fig:cnn-performance}.}
  \end{subfigure}
  \hspace{0.5em}
  \begin{subfigure}[b]{.55\textwidth}
    \centering
    \resizebox{\textwidth}{!}{\input{figures/demoe2c0.pgf}}
    \caption{\label{fig:fsmp} An example giving \\ $\hat{t}_0 - t_0=\SI{1.94}{ns}$, $\mathrm{RSS}=\SI{59.3}{mV^2}$, $D_\mathrm{w}=\SI{0.69}{ns}$.}
  \end{subfigure}
  \caption{\label{fig:fsmp-performance}Demonstration of FSMP with $\num[retain-unity-mantissa=false]{1e4}$ waveforms in~\subref{fig:fsmp-npe} and one waveform in~\subref{fig:fsmp} sampled from the same setup as figure~\ref{fig:method}.  FSMP reconstructs the waveform and charges flawlessly.}
\end{figure}
In terms of $D_\mathrm{w}$, figure~\ref{fig:fsmp-npe} shows that FSMP is on par with CNN in figure~\ref{fig:cnn-npe}.  Figure~\ref{fig:fsmp} is a perfect reconstruction example where the true and reconstructed charges and waveforms overlap.  Estimators for $t_0$ and $\mu$ in eq.~\eqref{eq:fsmpcharge} is an elegant interface to event reconstruction, eliminating the need of $\hat{t}_\mathrm{KL}$ and $\hat{\mu}_\mathrm{KL}$ in section~\ref{sec:pseudo}.  A low $\hat{t}_0 - t_0=\SI{1.94}{ns}$ aligns with the fact that $\hat{t}_0$ of eq.~\eqref{eq:fsmpcharge} is unbiased.  The superior performance of FSMP attributes to sparsity and positiveness of $q'_i$, correct modeling of $V_\mathrm{PE}$, $q'$ distribution and white noise.

\section{Summary and discussion}
\label{sec:discussion}

This section will address the burning question: which waveform analysis method should my experiment use?  We surveyed 8 methods, heuristics~(figure~\ref{fig:method}), deconvolution~(figure~\ref{fig:deconv}), neural network~(figure~\ref{fig:cnn-performance}) and regressions(figures~\ref{fig:dcf}--\ref{fig:fsmp-performance}), from the simplest to the most dedicated\footnote{The source codes are available on GitHub \url{https://github.com/heroxbd/waveform-analysis}.}.  To make a choice, we shall investigate the light curve reconstruction precision under different light intensities $\mu$.

\subsection{Performance}

\begin{figure}[H]
    \centering
    \resizebox{0.9\textwidth}{!}{\input{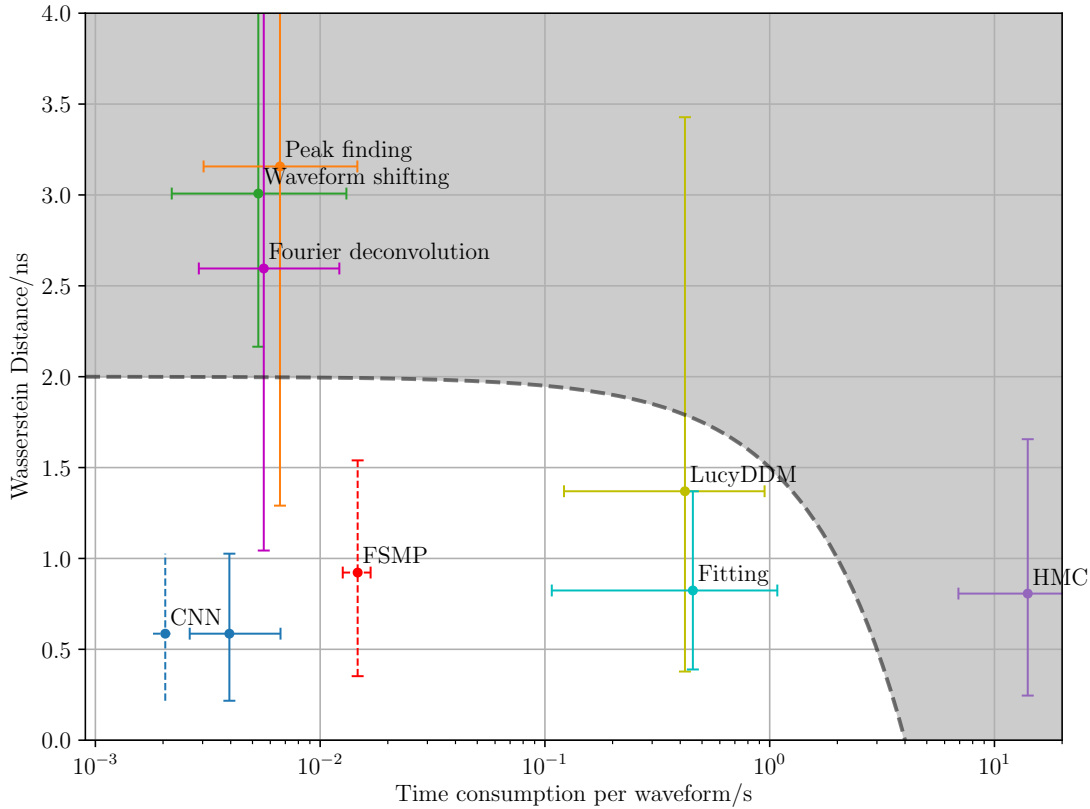}}
    \caption{\label{fig:chargesummary} Performance of algorithms in terms of $D_\mathrm{w}$ and time consumption, evaluated on the same dataset as figure~\ref{fig:method}. Central points are the average results of $\num[retain-unity-mantissa=false]{1e4}$ waveforms from specific $\mu$ values.  Error bars are 5--95 percentiles.  Fitting stands for direct charge fitting. The time consumed by Fitting, HMC and FSMP include the LucyDDM pre-conditioner's initialization time.  CNN's time consumption is measured for inference in two conditions: GPU\protect\footnotemark~(dashed error bars) and CPU\protect\footnotemark~(solid error bars).  Training a CNN is a one-time job, and its cost is not included in the plot.}
\end{figure}

We constrain the candidates by time consumption, algorithm category and $D_\mathrm{w}$.  Figure~\ref{fig:chargesummary} shows the $D_\mathrm{w}$ and time consumption summary of all eight methods with the same waveform sample as figure~\ref{fig:method}.
\addtocounter{footnote}{-2}
\stepcounter{footnote}\footnotetext{One NVIDIA\textsuperscript{\textregistered} A100 GPU (40GB PCIe). }
\stepcounter{footnote}\footnotetext{One CPU core of AMD EYPC\texttrademark\ 7742. }

The $D_\mathrm{w}$ performance of waveform shifting, peak finding and Fourier deconvolution are suboptimal.  Like CNN, they are the fastest because no iteration is involved.  Fitting has $\num{\sim 100}$ iterations, while LucyDDM and FSMP have $\num{\sim 1000}$ iterations, making them 1-2 orders of magnitudes slower.  HMC is too expansive and its principle is not too different from FSMP.  We shall focus on CNN, LucyDDM, Fitting and FSMP in the following.  

\begin{figure}[H]
  \begin{subfigure}[b]{\textwidth}
    \resizebox{\textwidth}{!}{\input{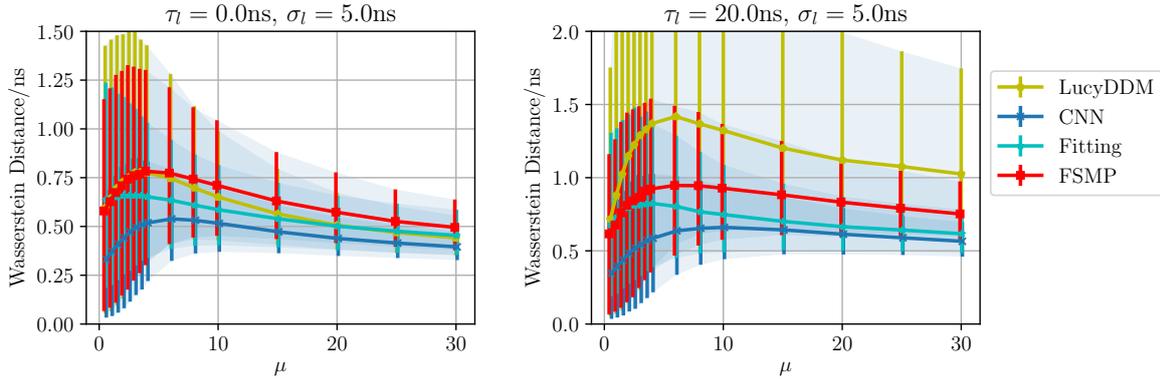}}
    \caption{\label{fig:wdistsummary}Dependence of Wasserstein distance on light intensity.}
  \end{subfigure}

  \vspace{0.5em}
  \begin{subfigure}[b]{\textwidth}
    \resizebox{\textwidth}{!}{\input{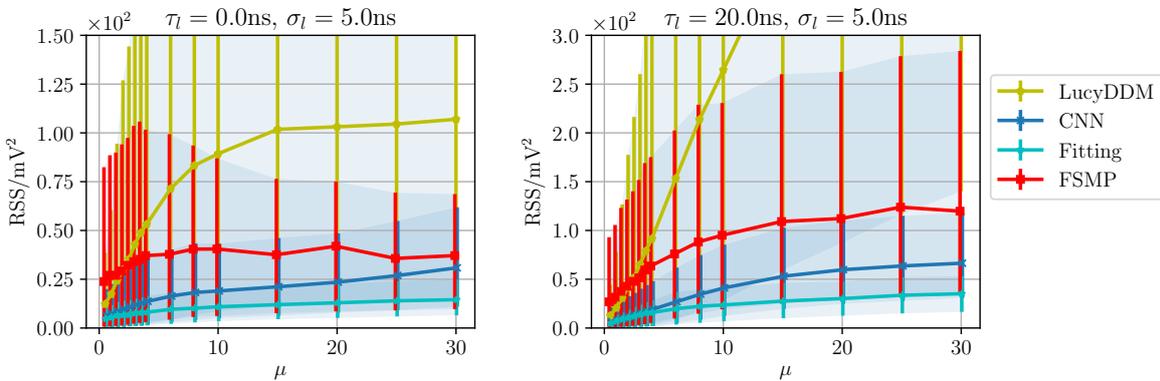}}
    \caption{\label{fig:rsssummary}Dependence of residual sum of squares on light intensity.}
  \end{subfigure}
  \caption{\label{fig:summary}The dependence of $D_\mathrm{w}$~\subref{fig:wdistsummary} and RSS~\subref{fig:rsssummary} on light intensity $\mu$ for typical Cherenkov (left) and scintillation (right) configurations.  Central points, error bars and method abbreviations have the same meaning as figure~\ref{fig:chargesummary}.  With more pile-ups, $D_\mathrm{w}$ tends to converge while RSS diverges.  The pile-up effect is more significant for the Cherenkov case because the time scale of the light curve is narrower. }
\end{figure}

The $D_\mathrm{w}$ and RSS dependence on $\mu$ of LucyDDM, Fitting, CNN and FSMP are plotted in figures~\ref{fig:wdistsummary} and \ref{fig:rsssummary}.  When $\mu$ increases the $D_\mathrm{w}$ of different methods approach each other, while the RSS diverges.  Notice that in the qualitative discussion, large $N_\mathrm{PE}$, large light intensity $\mu$ and large pile-ups are used interchangeably.  The $D_\mathrm{w}$ decrease-before-increase behavior is observed in section~\ref{sec:cnn} that with large $N_\mathrm{PE}$ the overall PE times dominate.  It is harder to be good at $D_\mathrm{w}$ and RSS with larger $N_\mathrm{PE}$, but FSMP achieves the best balance.  The Markov chains of FSMP have room for efficiency tuning.  Furthermore, implementing it in field-programmable gate array~(FPGA) commonly found in front-end electronics will accelerate waveform analysis and reduce the volume of data acquisition.  It is also interesting whether a neural network can approximate FSMP.

CNN and Fitting are the kings of $D_\mathrm{w}$ and RSS, because their loss functions are chosen accordingly.  It is informative to study the $\hat{q}_i$ distribution that is not related to the loss function of any method.

\subsection{Charge fidelity and sparsity}
\label{sec:sparsity}

\begin{figure}[H]
  \centering
  \resizebox{0.6\textwidth}{!}{\input{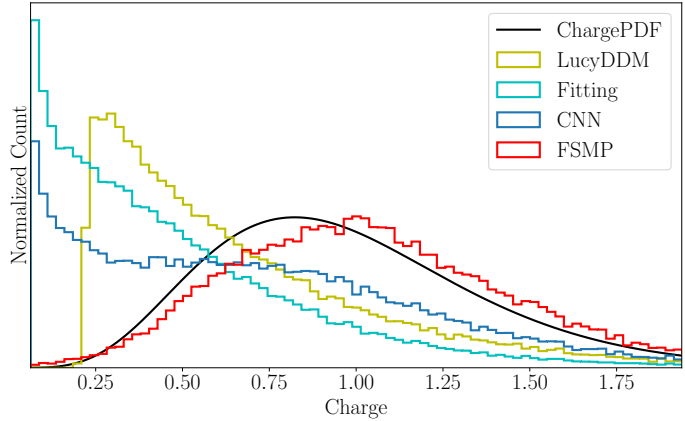}}
  \caption{\label{fig:recchargehist} $\hat{q}_i$ distributions on the same waveform dataset as figure~\ref{fig:method}.  Method abbreviations are defined in figure~\ref{fig:chargesummary}. ``ChargePDF'' is the charge distribution of simulation input in section~\ref{subsec:spe}. The cut-off near 0 in LucyDDM is an artifact of thresholding in eq.~\eqref{eq:fdconv2}.}
\end{figure}

All the discussed methods output $\hat{q}_i$ as the inferred charge of the PE at $t_i'$.  Evident in figure~\ref{fig:recchargehist}, FSMP retains the true charge distribution.  It is the only method modeling PE correctly.

In contrast, LucyDDM, Fitting and CNN distributions are severely distorted.  During the optimization process of $D_\mathrm{w}$ or RSS, $N_\mathrm{s}$ is a constant. Many $\hat{q}_i$ are inferred to be fragmented values.  Retaining charge distribution is a manifestation of sparsity.  FSMP has the best sparsity because it chooses a PE configuration $\bm{z}$ before fitting $\hat{q}_i$.  Since CNN is $D_\mathrm{w}$ orientated discussed in section~\ref{sec:cnn}, it is better than fitting, although the latter has self-regulated sparsity in theory.

For large $N_\mathrm{PE}$, the sparsity condition is by definition lost.  The equivalence of charge fidelity and sparsity implies that FSMP performs similarly to others for these cases, as we shall see in the following sections.

\subsection{Inference of incident light}
\label{subsec:timeresolution}

\begin{figure}[H]
  \begin{subfigure}[b]{\textwidth}
    \centering
    \resizebox{0.99\textwidth}{!}{\input{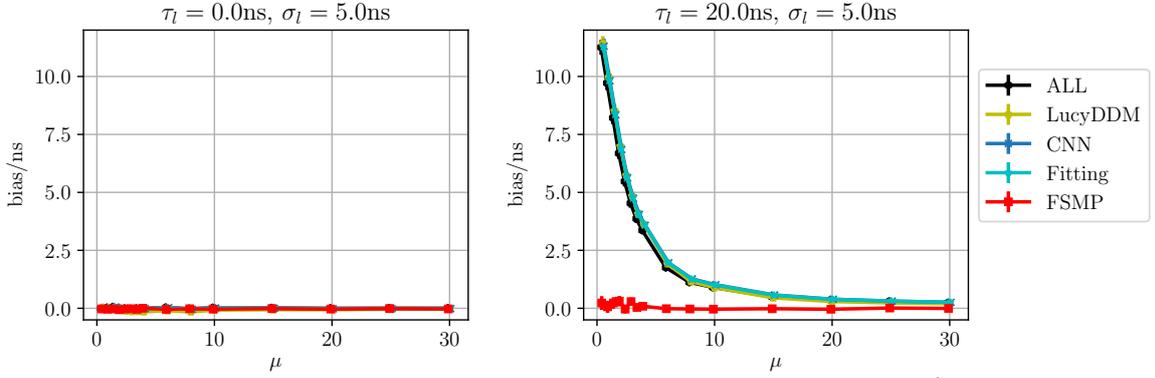}}
    \vspace{-0.5em}
    \caption{\label{fig:biasmethods} Sample average estimations of time-reconstruction biases $\E[\hat{t} - t_0]$.}
  \end{subfigure}

  \vspace{0.5em}
  \begin{subfigure}[b]{\textwidth}
    \centering
    \resizebox{0.99\textwidth}{!}{\input{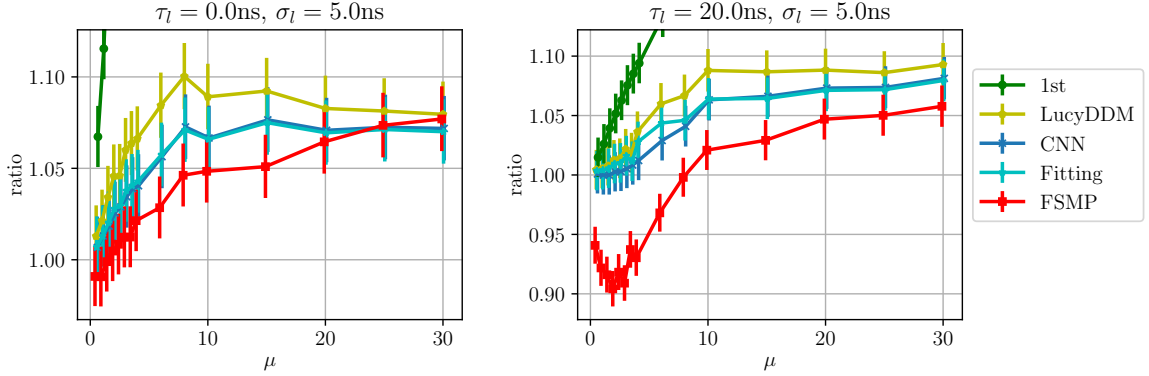}}
    \vspace{-0.5em}
    \caption{\label{fig:deltamethods} Sample variance estimations of time-resolution ratios $\sqrt{\frac{\Var[\hat{t} - t_0]}{\Var[\hat{t}_\mathrm{ALL} - t_0]}}$. ``1st'' is a reproduction of figure~\ref{fig:reso-diff}.}
  \end{subfigure}

  \vspace{0.5em}
  \begin{subfigure}[b]{\textwidth}
    \centering
    \resizebox{0.99\textwidth}{!}{\input{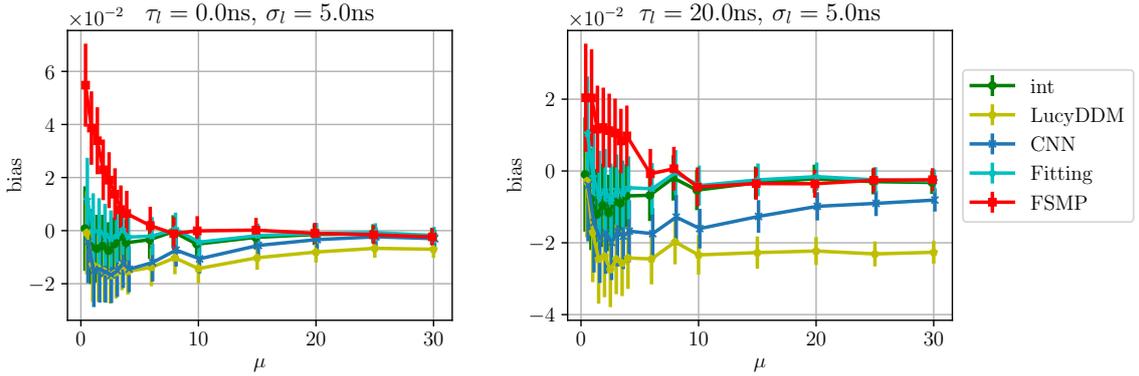}}
    \vspace{-0.5em}
    \caption{\label{fig:biasmu} Sample average estimation of intensity-reconstruction biases $\frac{\E[\hat{\mu} - \mu]}{\mu}$.}
  \end{subfigure}

  \vspace{0.5em}
  \begin{subfigure}[b]{\textwidth}
    \centering
    \resizebox{0.99\textwidth}{!}{\input{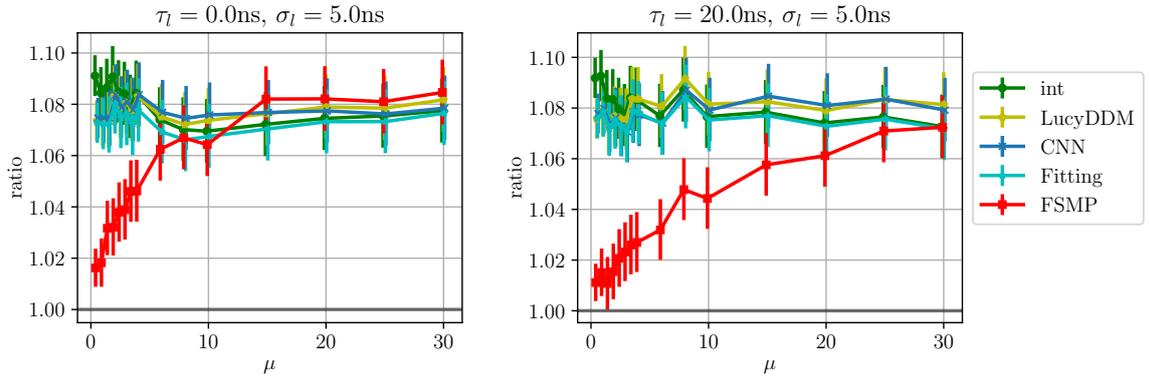}}
    \vspace{-0.5em}
    \caption{\label{fig:deltamu} Sample variance estimation of intensity-resolution ratios $\frac{\sqrt{\Var[\hat{\mu}]} / \E[\hat{\mu}]}{\sqrt{\Var[N_\mathrm{PE}]} / \E[N_\mathrm{PE}]}$.}
  \end{subfigure}
  \caption{Incident light analysis results for the two typical cases of Cherenkov~(left) and scintillation~(right).  ``ALL'' and ``1st'' are the $\hat{t}_\mathrm{ALL}$ estimator defined in eq.~\eqref{eq:2}. ``int'' is the $\hat{\mu}_Q$ by eq.~\eqref{eq:mu-q}.  LucyDDM, Fitting, CNN use eqs.~\eqref{eq:pseudo} and \eqref{eq:pseudo-mu}.  FSMP has its own natural $\hat{t}_0$ and $\hat{\mu}$ estimators in eq.~\eqref{eq:fsmpcharge}. Error bars are 5--95 percentiles calculated from $t$~(figures~\ref{fig:biasmethods} and \ref{fig:biasmu}), $F$~\subref{fig:deltamethods} and $\chi^2$~\subref{fig:deltamu} statistics.}
\end{figure}

In figure~\ref{fig:summary}, we show the dependence on $\mu$ of bias~(figure~\ref{fig:biasmethods}) and resolution~(figure~\ref{fig:deltamethods}) for different time estimators in the two typical experimental setups.  From figure~\ref{fig:biasmethods}, we see that the $t_0$ estimation biases are all similar to that of $\hat{t}_\mathrm{ALL}$.  In the right of figure~\ref{fig:biasmethods}, the biases of LucyDDM, Fitting and CNN for the scintillation configuration at small $\mu$ are intrinsic in exponential-distributed MLEs.  Conversely, the FSMP $t_0$ estimator of eq.~\eqref{eq:fsmpcharge} is unbiased by constructing from samples in the Markov chain.

People often argue from difficulties for large pile-ups that waveform analysis is unnecessary.  Comparing figures~\ref{fig:reso-diff} and \ref{fig:deltamethods}, it is a myth.  Although $\hat{t}_\mathrm{1st}$ is more precise for large light intensity, all the waveform analysis methods provide magnificently better time resolutions than $\hat{t}_\mathrm{1st}$, more than twice for $\mu>20$ in Cherenkov setup.  FSMP gives the most significant boost.  Such improvement in time resolution amounts to the position resolution, which benefits fiducial volume, exposure and position-dependent energy bias.

The message is clear from figure~\ref{fig:deltamethods}: any PMT-based experiment that relies on time with PMT occupancy $\mu$ larger than 3 should employ waveform analysis.

In contrast to time, inference of light intensity uses empty waveforms as well.  We append the waveform samples by $10^4 \times e^{-\mu} / (1-e^{-\mu})$ empty ones.  The number is proportional to the Poisson prediction.  It is equivalent to appending the same amount of zeros to the $\hat{\mu}$'s. The QDC integration estimator $\hat{\mu}_Q$~(``int'' in figures~\ref{fig:biasmu} and~\ref{fig:deltamu}) is ubiquitous and is plotted together with the four waveform analysis methods.

In figure~\ref{fig:biasmu}, the biases of $\hat{\mu}$ of the four methods are within \SI{6}{\percent} and disappear for large $\mu$ expect LucyDDM.  The tendency of LucyDDM comes from the thresholding and scaling in eq.~\eqref{eq:fdconv2}.  For low $\mu$, the upward bias of FSMP and Fitting is due to PE granularity.  The charge $q$ of one PE can fluctuate close to 2 or 0, but eqs.~\eqref{eq:gd-q} and \eqref{eq:fsmpcharge} favor 2 more than 0 in waveforms.  We shall leave the amendment of the bias to event reconstruction in our subsequent publications.

For large $\mu$ the four methods are similar in intensity resolution to $\hat{\mu}_Q$~(figure~\ref{fig:deltamu}).  The resolution ratios of them all approach $1.08 = \sqrt{1 + 0.4^2}$, consistent with eq.~\eqref{eq:energy} if white noise is ignored.  For small $\mu$, FSMP gives the best resolution by correctly modeling charge distributions, as predicted in figure~\ref{fig:recchargehist}.  Like the hit estimator $\hat{\mu}_\mathrm{hit}$ in section~\ref{sec:intensity-mu}, it eliminates the influence of $\Var[q]$ and $\Var[\epsilon]$ in eq.~\eqref{eq:energy}.  But unlike $\hat{\mu}_\mathrm{hit}$, FSMP also works well for a few PEs.  More importantly, it provides a smooth transition from the photon-counting mode to the analog mode with the best intensity resolution of all.   In the scintillation case of figure~\ref{fig:deltamu}, FSMP approaches the resolution lower bound $\Var[N_\mathrm{PE}]$ set by the PE truths for $\mu < 5$, which is the ultimate waveform analysis in that we can hardly do any better.

In the fluid-based neutrino and dark matter experiments, $\mu < 5$ is the sweet spot for \si{MeV} and \si{keV} physics respectively.  The intensity resolution boost in figure~\ref{fig:deltamu} converts directly into energy resolution.  It varies depending on PMT models for different $\Var[q]$ and $\Var[\epsilon]$.  In our scintillation setup, the improvement is up to $\times 1.07$~(figure~\ref{fig:deltamu}) . Good waveform analysis has the potential to accelerate discovery and broaden the physics reach of existing detectors. 
\section{Conclusion}
\label{sec:conclusion}

We develop and compare a collection of waveform analysis methods.  We show for $N_\mathrm{PE} > 1$, better time resolution can be achieved than the first PE hit time.  FSMP gives the best accuracy in both time and intensity measurements, while CNN, Fitting and LucyDDM follow.  We find significant time~($\times 2$) and intensity~($\times 1.07$) resolution improvements at higher and lower light intensities, respectively.    FSMP opens the opportunity to boost energy resolution~($\times 1.07$) and particle identification in PMT-based neutrino and dark matter experiments.  We invite practitioners to deploy FSMP for the best use of the data and tune it for faster execution.

\acknowledgments
We are grateful to the participants and organizers of the \textit{Ghost Hunter 2019} online data contest.  CNN, Fitting and LucyDDM in this article are developed from the ones submitted to the contest.  We would like to thank Changxu Wei and Wentai Luo for sharing findings on direct charge fitting.  The idea of multivariate Gaussian emerged during a consulting session at the Center for Statistical Science at Tsinghua University.  The authors also want to thank Professor Kai Uwe Martens for discussions on PE granularity and the anonymous referee for providing the constructive comments.  The corresponding author would like to express deep gratitude to KamLAND, XMASS and JUNO collaborations for nurturing atmosphere and encouraging discussions on waveform analysis.  This work is supported by the undergraduate innovative entrepreneurship training programme of Tsinghua University, in part by the National Natural Science Foundation of China (12127808, 11620101004), the Key Laboratory of Particle \& Radiation Imaging (Tsinghua University). 


\bibliographystyle{JHEP}
\bibliography{ref}

\end{document}